\documentclass[article]{elsarticle}
\usepackage[margin=2.5cm]{geometry}
\usepackage{lineno,hyperref}
\modulolinenumbers[5]
\pdfoutput=1 

\usepackage[utf8]{inputenc}
\journal{.}
\usepackage{amsmath,color}
\numberwithin{equation}{section}

\def\nonu{\nonumber\\}

\def\t #1{\tau_{#1}}
\def\non{\nonumber\\}

\def\e{\,{\rm e}}

\def\veps#1{\varepsilon_{#1}}

\def\kk#1#2{k_{#1}\cdot k_{#2}}

\def\veps{\varepsilon}

\def\tanh{\rm tanh}
\def\half{{1\over 2}}

\def\Z{{\mathchoice {\hbox{$\sf\textstyle Z\kern-0.4em Z$}}
{\hbox{$\sf\textstyle Z\kern-0.4em Z$}}
{\hbox{$\sf\scriptstyle Z\kern-0.3em Z$}}
{\hbox{$\sf\scriptscriptstyle Z\kern-0.2em Z$}}}}
\def\abs#1{\left| #1\right|}

        \def\slash#1{#1\!\!\!\raise.15ex\hbox {/}}
\newcommand{\slD}{\,\raise.15ex\hbox{$/$}\kern-.27em\hbox{$\!\!\!D$}}
\newcommand{\slpartial}{\raise.15ex\hbox{$/$}\kern-.57em\hbox{$\partial$}}

\def\epsk#1#2{\varepsilon_{#1}\cdot k_{#2}}
\def\epseps#1#2{\varepsilon_{#1}\cdot\varepsilon_{#2}}
\def\Gd{\dot{G}}
\def\Gdd{\ddot{G}}

\def\no{\noindent}

\def\kinb{{1\over 4}\dot x^2}

\def\be{\begin{equation}}
\def\ee{\end{equation}\noindent}
\def\bear{\begin{eqnarray}}
\def\ear{\end{eqnarray}\noindent}
\def\bec{\blue\begin{equation}}
\def\eec{\end{equation}\black\noindent}
\def\bearc{\blue\begin{eqnarray}}
\def\earc{\end{eqnarray}\black\noindent}
\def\benn{\begin{enumerate}}
\def\enn{\end{enumerate}}
\def\ee{&=&}

\def\mn{{\mu\nu}}

\def\tr{{\rm tr}\,}

\def\e{\,{\rm e}}

\def\b0{{\bf 0}}

\def\G#1#2{G_{#1#2}}

\def\4piTD{{(4\pi T)}^{-{D\over 2}}}
\def\4piT4{{(4\pi T)}^{-2}}

\def\edot{\,\e^{(\cdot)}} 

\begin{document}

\begin{frontmatter}

\title{The QED four -- photon amplitudes off-shell: part 1}

\author[a,b]{Naser Ahmadiniaz}
\ead{n.ahmadiniaz@hzdr.de}
\author[c]{Cristhiam Lopez-Arcos}
\ead{cmlopeza@unal.edu.co}
\author[b,d]{Misha A. Lopez-Lopez}
\ead{misha.lopez@umich.mx}
\author[d]{Christian Schubert\corref{correspondingauthor}}
\cortext[correspondingauthor]{Corresponding author}
\ead{christianschubert137@gmail.com}

\address[a]{Helmholtz-Zentrum Dresden-Rossendorf, Bautzner Landstra\ss e 400, 01328 Dresden, Germany}
\address[b]{Center for Relativistic Laser Science, Institute for Basic Science, 61005 Gwangju, Korea}
\address[c]{Escuela de Matem\'{a}ticas, Universidad Nacional de Colombia Sede Medell\'{i}n, Carrera 65 $\#$ 59A--110, Medell\'{i}n, Colombia}
\address[d]{Instituto de F\'{i}sica y Matem\'{a}ticas
Universidad Michoacana de San Nicol\'{a}s de Hidalgo
Edificio C-3, Apdo. Postal 2-82
C.P. 58040, Morelia, Michoac\'{a}n, M\'{e}xico}

\begin{abstract}
The present paper is the first in a series of four where we use the worldline formalism to obtain the QED four-photon amplitude completely off-shell. We present the result explicitly in terms of hypergeometric functions, and derivatives thereof, for both scalar and spinor QED. The formalism allows us to unify the scalar and spinor loop calculations, avoiding the usual breaking up of the amplitude into Feynman diagrams, and to achieve manifest transversality at the integrand level as well as UV finiteness term by term by an optimized version of the integration-by-parts procedure originally introduced by Bern and Kosower for gluon amplitudes. The full permutation symmetry is maintained throughout, and the amplitudes get projected naturally into the basis of five tensors introduced by Costantini et al. in 1971. Since in many applications of the ``four-photon box'' some of the photons can be taken in the low-energy limit, and the formalism makes it easy to integrate out any such leg, apart from the case of general kinematics (part 4) we also treat the special cases of one (part 3) or two (part 2) photons taken at low energy. In this first part of the series, we summarize the application of the worldline formalism to the N-photon amplitudes and its relation to Feynman diagrams, derive the optimized tensor-decomposed integrands of the four-photon amplitudes in scalar and spinor QED, and outline the computational strategy to be followed in parts 2 to 4. We also give an overview of the applications of the four-photon amplitudes, with an emphasis on processes that involve some off-shell photons. The case where all photons are taken at low energy (the ``Euler-Heisenberg approximation'') is simple enough to be doable for arbitrary photon numbers, and we include it here for completeness.
\end{abstract}


\end{frontmatter}


\tableofcontents

	
\section{Introduction: Photon amplitudes in QED}\label{sec:intro}
	
	Classical electrodynamics is described by Maxwell's equations, which are linear in both sources and fields.  Non-linear corrections to Maxwell's theory appear as purely quantum effects that violate the 
	superposition principle and, among other things, give rise to the possible polarization of the vacuum.
	
\begin{figure}[h]
		\centering 
		\includegraphics[scale=0.85]{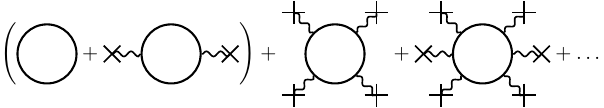}
		\caption{\label{eh} Diagrammatic perturbative expansion of the EH effective action, eq. (\ref{EH}). The first and second diagrams are show in parenthesis because they get
		 removed by on-shell renormalization. 
		 }
	\end{figure}

	Dirac's prediction of the positron \cite{Dirac1928-610,Dirac1931-60} led to the possibility of creation and annihilation of virtual particles, non-linear effects that have as a particular consequence the possibility of the light-by-light (LBL) scattering process. 
This effect was first studied in 1936 by W. Heisenberg and H. Euler \cite{Heisenberg1936-714} 
where they, still using Dirac's hole theory, obtained their famous closed-form expression one-loop effective Lagrangian 
for spinor particles in a constant background field. 
In modern conventions, the Euler-Heisenberg (EH) Lagrangian is given by 
	
	\begin{equation}
	\mathcal{L}^{}_{\rm EH}=-\frac{1}{8\pi^2}\int_0^\infty \frac{dT}{T^3}{\rm e}^{-m^2T} \left\{\frac{(e a T)(e b T)}{\tan(e a T)\tanh ({\it ebT})} -\frac{2}{3}(eT)^2 \mathcal{F}-1 \right\} ,
	\label{EH}
	\end{equation}
with
	\begin{equation}
	a=\left(\sqrt{\mathcal{F}^2+\mathcal{G}^2}-\mathcal{F} \right)^{1/2}~,~~~~~~~ b=\left(\sqrt{\mathcal{F}^2+ \mathcal{G}^2}+\mathcal{F}\right)^{1/2},
	\end{equation}
 with the two invariants of the Maxwell field
	\begin{eqnarray}
	-2\mathcal{F}=-\frac{1}{2}F_{\mu\nu}F^{\mu\nu}&=& \vec{E}^2-\vec{B}^2\,,\nonumber\\
	-\mathcal{G}=-\frac{1}{4}F_{\mu\nu}\tilde{F}^{\mu\nu}& =&\vec{E}\cdot\vec{B}\, .\nonumber\\
	\end{eqnarray}
	Shortly later, 
	an analogous expression was found for the scalar QED by Weisskopf \cite{Weisskopf1936-1},
	\begin{equation}
	\mathcal{L}^{}_{\rm W}=\frac{1}{16\pi^2}\int_0^\infty \frac{dT}{T^3}{\rm e}^{-m^2T} \left\{\frac{(e a T)(e b T)}{\sin(e a T)\sinh ({\it ebT})} +\frac{1}{6}(eT)^2 \mathcal{F}-1 \right\}.
	\label{weisskopf}
	\end{equation}
	See \cite{Dunne2005-445} for a comprehensive review of the effective Lagrangians (\ref{EH}) and \eqref{weisskopf}, their
	applications and generalizations. 
	They are non-perturbative in the field, but have a perturbative expansion in powers of the field (``weak-field expansion''). For the spinor QED case, it corresponds to the Feynman diagrams shown in Fig. \ref{eh},
	where the ``cross'' denotes the interaction with the field.
	For scalar QED, there are additional diagrams involving the quartic seagull vertex. 
	When specialized to a multi-photon background, these diagrams (now without the crosses) 
	represent the $N$-photon amplitudes in the low-energy limit, where all the photon energies are small compared to the electron mass. 

\bigskip

	\bigskip

	
	The order $e^4$ (or $\alpha^2$, $\alpha = \frac{e^2}{4\pi}$ being the fine structure constant)
	of this expansion gives the lowest-order 
	photon-photon scattering diagram, or rather diagrams, since one has to sum over
	the six possible orderings of the four photons (see below, Fig. \ref{fig-photonphoton}). 
		
In this low-energy limit, the first calculation of the cross section for photon-photon scattering was done by H. Euler and B. Kockel \cite{Euler1935-246,Euler1936-398}. Shortly later, this was followed by a calculation of the opposite high-energy limit by A. Akhiezer et al.  \cite{Akhiezer1936vzu,Achieser1937ywd}. 
	 
	In the low-energy limit, it is a textbook exercise to obtain an explicit expression for the four-photon amplitude (for spinor QED) from the effective Lagrangian \eqref{EH} (see, e.g., \cite{Itzykson1980rh,Dittrich2000zu}). Introducing the field strength tensor 
	$f_i^{\mu\nu}$ for photon $i$, 
	
	\begin{equation}\label{fmunu}
	f_i^{\mu\nu} \equiv k_i^\mu \veps_i^\nu- \veps_i^\mu k_i^{\nu},
	\end{equation}
	related to the field strength tensor $F^\mn$ by
	\begin{equation}
	F^{\mu\nu} = i f_{\rm tot}^{\mu\nu} \equiv  i \sum_{i=1}^N f_i^{\mu\nu},
	\end{equation}
	the spinor amplitude for low energy photons can be written as
	\bear
	\label{M}
	{\cal M} &=& \frac{2  \alpha^2}{45m^{4}}  \Big\{ 14 \left[ {\rm tr}(f_1f_2f_3f_4) + {\rm tr}(f_1f_2f_4f_3) +{\rm tr}(f_1f_3f_2f_4) \right]\non
	&~& - 5 \left[ {\rm tr}(f_1f_2) {\rm tr}(f_3f_4) + {\rm tr}(f_1f_3) {\rm tr}(f_2f_4) + {\rm tr}(f_1f_4) {\rm tr}(f_2f_3) \right]\Big\}\,.
	\ear
	Then, in the center of mass frame and after some algebra one can obtain the cross section 
	\begin{equation}\label{lowcross}
	\sigma = \frac{973}{10125} \left(\frac{\alpha^4}{\pi m^{2}}\right) \left(\frac{\omega}{m}\right)^6  ~~~~~{\rm for}~~~~~~ \frac{\omega}{m} \ll 1 .
	\end{equation}
	Similarly, one can use the Weisskopf Lagrangian \eqref{weisskopf} to compute the same cross section for scalar QED. 
	
	On the other hand, computing the cross section for LBL scattering in the high energy limit is a much harder task. Even so, Akhiezer et al. were able to obtain the following formula for the differential cross section in this limit \cite{Akhiezer1936vzu,Achieser1937ywd}
	\begin{equation}
	\sigma = \frac{e^8}{2\pi \omega^2}~ C ~~~~~{\rm for}~~~~~~ \frac{\omega}{m} \gg 1, ~~~m=0\,,
	\end{equation}
	but with a numerical constant $C$ that they were unable to calculate. 
	They also showed that $\sigma$ has a maximum in the region $\omega \sim m$. 
	A numerical value for the constant $C$ was given in \cite{Berestetskii1982-501} as $\frac{C}{2\pi}=4.7$, and an analytical one finally in \cite{Akhiezer1981}
	\begin{equation}\label{constantC}
	C = \frac{108}{5} + \frac{13}{2}\pi^2 - 8\pi^2 \zeta(3) + \frac{148}{225}\pi^4 - 24\zeta(5).
	\end{equation}
	For a comprehensive review of the early history of these calculations see \cite{Scharnhorst2017photonphoton}. 
	
\begin{figure}[h]
		\centering 
		\includegraphics[width=.7\textwidth]{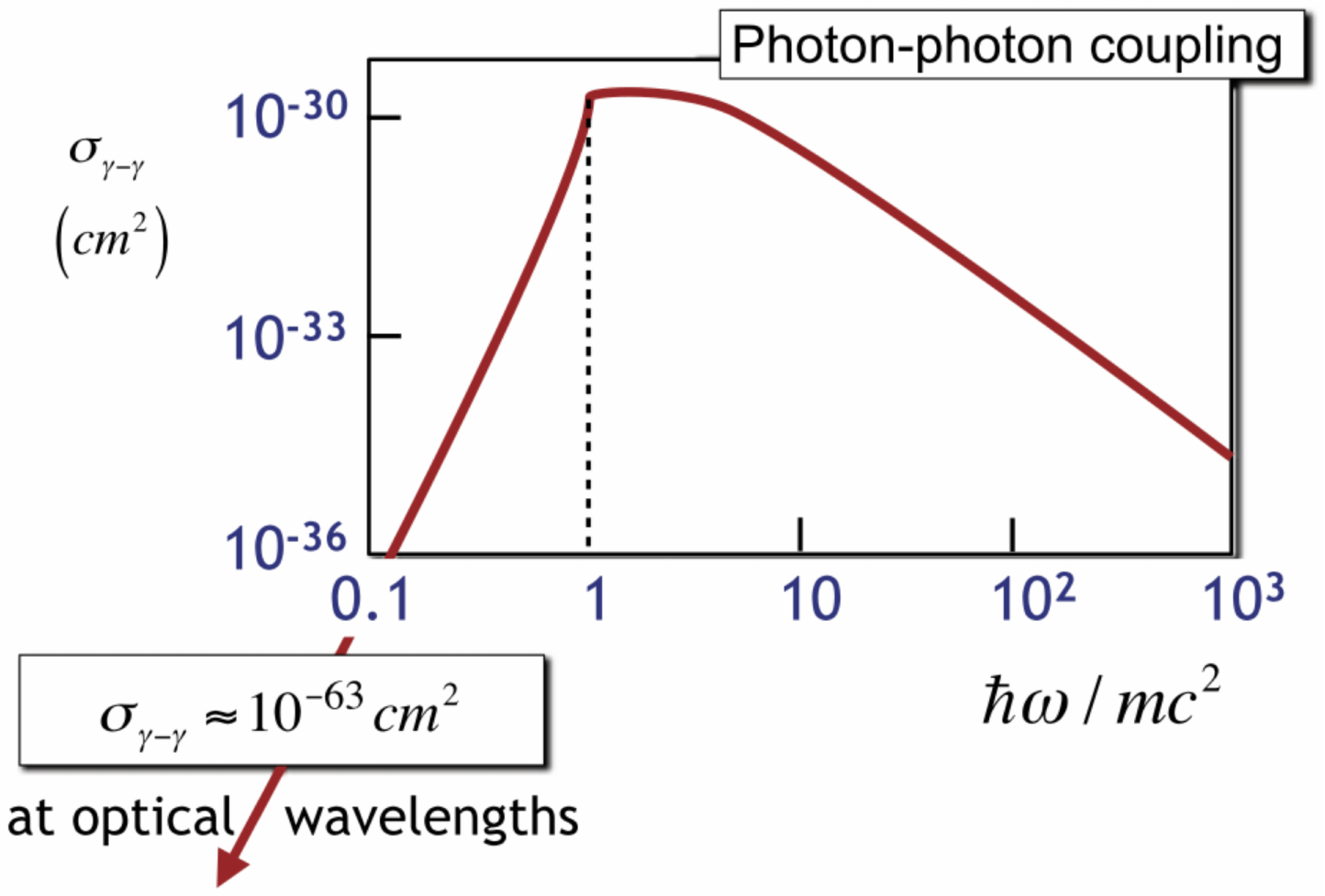}
		\caption{\label{grapcrossection} Logarithmic plot of the LBL cross section in units of the electron mass. This graph exhibit a maximum in $\omega \sim m$, as predicted in \cite{Akhiezer1936vzu}. The cross section is extremely small, in particular at optical wavelengths. Adapted from \cite{Tommasini2014-137}. }
	\end{figure}

	The first treatment of the photon-photon scattering amplitude for arbitrary 
	on-shell kinematics was done by Karplus and Neuman in 1950 \cite{Karplus1950-380,Karplus1951-776}.
	They analyzed the tensor structure of the four-photon amplitude, 
	showed its finiteness and gauge invariance, and managed to express it by
	three-parameter integrals of rational functions of the external momenta. 
	In the mid-sixties De Tollis \cite{DeTollis:1964una,DeTollis:1965vna} recalculated the amplitude using dispersion relation techniques, which led to a more compact form of the result. Using the results of either Karplus and Neuman or De Tollis, it is possible to compute and plot the cross section (see Fig. \ref{grapcrossection}), and to compute the constant $C$ in (\ref{constantC}).

From a computational point of view, this essentially settled the on-shell case. 
Further work on on-shell photonic amplitudes focused on 
higher-point and higher-loop amplitudes, and generalization to SUSY QED and electroweak contributions. 
The six photon amplitude was computed in massless scalar QED \cite{Bernicot_2008} and the $N$-photon amplitudes
in the low-energy limit for both scalar and spinor QED \cite{Martin2003-335,Edwards2018-198}.
For two-loop corrections to LBL scattering in spinor QED see \cite{Bern_2001}, in supersymmetric QED \cite{Binoth_2002}.
Photon-photon scattering via a W-boson loop has been studied in \cite{Jikia1994-453,Gounaris:1999gh}.

However, eventually the need was felt to calculate the four-photon amplitudes also with some legs off-shell.
Photons emitted or absorbed by an external field in general can not be assumed to obey on-shell conditions.
An important example is Delbr\"uck scattering, the elastic scattering of a photon by a nuclear electromagnetic field, where
the scattered photon can be taken real but the interaction with the field is described by virtual photons 
(see the left-hand diagram of Fig. \ref{fig-delbruck} below). 
This process, and similar ones involving interactions with the Coulomb field, prompted 
V. Costantini et al. in 1971 \cite{Costantini:1971cj} (see also \cite{Leo:1975fb})
to study the four-photon amplitude 
with two photons on-shell and two off-shell. 
In the same work, they performed a general analysis of the tensor structure of the four-photon amplitude, 	which from the beginning is written in terms of $138$ tensors. 
Using the permutation symmetry and the Ward identity they were able to reduce this number to a basis of five invariant tensors. 

\begin{figure}[h]
		\centering 
		\includegraphics[width=.4\textwidth]{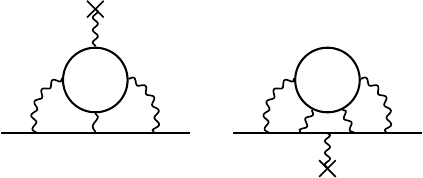}
		\caption{ Feynman diagrams contributing to the QED $g-2$ factor involving the one-loop
		four - photon process as a subdiagram.}
		\label{fig-g23loop}
	\end{figure}

Although their work dates back half a century, to the best of our knowledge no further progress has been
made towards the actual calculation of the fully off-shell case. 
In our opinion, this state of affairs is quite unsatisfactory considering the fact 
that multiloop QED calculations have long ago reached the stage where the LBL scattering diagram appears  
fully off-shell, for example, as a subdiagram in the calculation of the three- and higher loop anomalous magnetic momentum, see
Fig. \ref{fig-g23loop}.     

Thus the present series of four papers is devoted to the study of the fully off-shell four-photon amplitudes in scalar and spinor QED.
While this would be feasible using standard techniques, here we will take advantage of the recently developed
``string-inspired worldline formalism'' that has been shown to have significant technical advantages particularly in the computation
of photonic processes. 

Methods for amplitude calculations inspired by string theory emerged in the seventies with the work of 
J.L. Gervais and A. Neveu \cite{Gervais1972381} where they observed some simplification of quantum field theory computations in the limit of infinite string tension. In the nineties, it led Z. Bern and D. Kosower to start a systematic investigation of the field theory limit at tree- and loop-level using various string models \cite{Bern1991-1669}. In particular, they 
established a new set of rules for the construction of the one-loop $N$-gluon amplitudes in QCD 
which bypasses many steps of the ordinary quantum field theory computation based on Feynman rules. 
The relation of these rules to the Feynman ones was clarified in \cite{Bern1992562}.
Later, M. Strassler \cite{Strassler1993} was able to rederive the same set of rules inside field theory, using representations
of the one-loop effective actions in terms of first-quantized, relativistic worldline path integrals and 
just some ideas and techniques from string theory. Such ``worldline path-integral representations'' had been obtained for
QED by Feynman in 1950/1 \cite{Feynman:1950ir,Feynman:1951gn}, but their computational advantages have been
recognized only in recent years following developments in string theory. While the work of Bern and Kosower had
focused on the non-abelian case, Strassler in \cite{Strassler:1992nc} analyzed the QED photon amplitudes and effective action, and
noted that the formalism allowed him, using certain integration by parts (IBP) that homogeneized the integrand and led to the
automatic appearance of photon field strength tensors, to arrive at an extremely compact integral
representation for the four-photon amplitudes in scalar and spinor QED. The IBP procedure was improved in
\cite{Schubert:1997ph} and \cite{Ahmadiniaz2013-132}, and in the present work we will refine it further to arrive at what we believe is the simplest possible integrand achievable for these amplitudes in the worldline formalism. 
Remarkably, it involves precisely the same tensor basis that in \cite{Costantini:1971cj} had been found using gauge invariance and symmetry considerations. 
The formalism makes it also easy to integrate out any leg
that is taken in the low-energy limit, and we take advantage of this to treat the off-shell four-photon amplitudes not only
for general kinematics, but also with one, two, or four legs taken at low energy 
(the case of just three low-energy legs would not not make sense because of energy-momentum conservation). 

In the present first part, we start with giving an overview over the manifold uses of the on-shell and off-shell 
four-photon amplitudes in section two.
In section three, we describe the application of the worldline formalism to the QED photon amplitudes in general, and 
in the following section zoom in on the four-photon case.  After contrasting the formalism with the standard Feynman one
in section five, section six is devoted to the problem of constructing a minimal tensor basis for the four-photon amplitudes. 
In section seven we obtain some ``matching identities'' by comparing the integrands of the four-photon amplitudes before 
and after the integration-by-parts procedure. In section eight we settle the case where all photons are of low
energy, which is easy enough to be doable explicitly even for an arbitrary number of photons. 
In section nine we outline the computational strategies to be followed in the remaining "two-low", "one-low" and
"zero-low" cases, to be treated in part 2, 3 and 4, respectively. 	
In section ten, we summarize our approach and point out further possible ramifications. 
	
\section{Phenomenological uses of the QED four-photon amplitude}

\begin{figure}[h]
		\centering 
		\includegraphics[width=.45\textwidth]{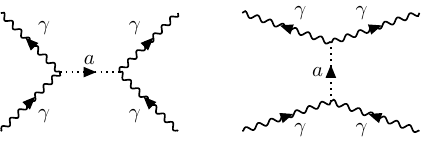}
		\caption{Photon-photon scattering induced by an axion-like particle.}
		\label{fig-alp}
\end{figure}

Unfortunately, the cross section 
according to equation (\ref{lowcross}) behaves like $\omega^6$ at small photon energies, 
which so far has prevented detection in the optical range \cite{Bernard2000,Bernard2006}, 
although the development of new high intensity lasers has recently give rise to new proposals of laser experiments \cite{Lundin2006-043821,Tommasini2009-043,Gies2018-036022,Gies2018-076002,Moulin:2002ya,Gies:2022pla,Gies2021lbl} for the direct measurement of LBL scattering (see also \cite{Fedotov:2022ely} for a contemporary review of QED with intense external fields). 
In a somewhat indirect manner, the photon-photon cross section was finally observed in 2017 at GeV energies
in ultra-peripheral heavy-ion collisions at LHC by the ATLAS collaboration
\cite{ATLAS2017-852,Vysotsky:2018pgf,CMS2019fhl,ATlAS2019-052001,Schoeffel:2020svx,Krintiras:2022jxa}.
There is an alternative way by which LBL interactions can be studied using relativistic heavy-ion collision \cite{DEnterria2013-080405,Klusek-Gawenda2016-044907,Ellis:2022uxv}. The electromagnetic field strengths of relativistic ions scale with the proton number ($Z$). For example, for a lead nucleus (Pb) with $Z=82$ 
the field can be up to $10^{25}~ {\rm V/m}$, much larger than the Schwinger limit ($E_c\propto 10^{18} V/m$) \cite{Sauter1931-742} above which QED corrections become important.	
Since the computation of the QED contribution to the photon-photon cross section relies only on the QED tree-level vertex
and the principles of QFT, its existence was, of course, hardly to be doubted.  
The main purpose of its measurement is thus the exclusion of new physics, that may manifest itself 
through fundamental or effective non-linear additions to the QED Lagrangian, a subject with a long history 
going back to the non-linear generalization of classical Maxwell theory proposed by Born-Infeld in 1934 \cite{Born:1934gh}.
See \cite{Trejo2011,Davila:2013wba} for a computation of the photon-photon scattering cross section combining the effect of the tree-level Born-Infeld term with the QED one-loop process, including the interference term. The above measurement by the ATLAS collaboration was used by
Ellis et. al \cite{Ellis2017-261802} to restrict the range of the free parameter appearing in the Born-Infeld Lagrangian. 
Probably the most-studied Beyond-the-Standard-Model (BSM) process inducing an effective four-photon vertex is the 
one shown in Fig. \ref{fig-alp}, involving the exchange of an axion-like particle (ALP). 

For light ALPs, this process might be measurable at laser energies (see \cite{Inada:2017lop,Shakeri:2020sin} and refs. therein), 
for heavy ALPs 
at future photon-photon colliders (see \cite{Inan:2020kif} and refs. therein). 
Photon-photon colliders are an old project that seems very promising for the understanding of new physics \cite{Jikia1994-453,Badelek2004-5097,Takahashi2018-893,Telnov2020gammagamma,Sangal2021}, but unfortunately their construction face many difficulties (the planned CLIC collider may be operated as a photon-photon collider \cite{Inan:2020kif}). 

Since in the present series of papers we are focussing on the off-shell case, 
let us now attempt to give a rough classification of the amplitudes and processes where the QED four-photon amplitude 
enters with some or all photons off-shell, proceeding from pure QED to the standard model and to extensions thereof. 

\subsection{Pure QED}

For QED processes, we can distinguish between purely photonic ones, and those involving also an external electric 
field (often of Coulomb type), an external magnetic field (usually treated as a constant field), or both.  

\subsubsection{Purely photonic processes}

The four-photon amplitude fully off-shell with general kinematics appears as a subdiagram in the calculation of the electron (or muon) anomalous magnetic moment starting from the four-loop level (right-hand side of Fig. \ref{fig-g23loop}). Since the contributions
of this type of diagrams turns out to be substantial, existing 
experimental results on the electron and muon $g-2$ \cite{Hanneke2008-120801,Muong-22006-072003} can be seen as
a test of the off-shell box, too. 

It should be observed that
in such calculations the sum over the three inequivalent permutations of the four photons takes a highly nontrivial character,
including the cancellation of the spurious UV divergences between them. 

Similarly the off-shell LBL box contributes as subdiagram
to the photon propagator, and thus to the QED beta function, starting from five loops. 

\begin{figure}[h]
		\centering 
		\includegraphics[width=.6\textwidth]{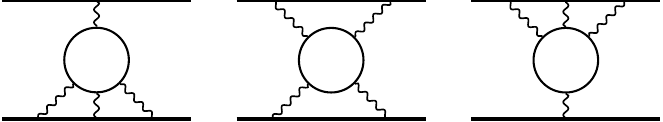}
		\caption{LBL Feynman diagrams contributing to the Lamb shift, with the thick line representing the nucleus and the thin one the electron.}
		\label{fig-bs}
\end{figure}

High-energy photon scattering close to the forward direction is an interesting case of a QED process where the higher-order
two-pair creation process $\gamma + \gamma \to e^+ + e^-+ e^+ + e^- $ dominates over the lower-order
$\gamma + \gamma \to e^+ + e^-$ one, for sufficiently large photon energies
\cite{Cheng:1970ef,Dittrich:1974xk,Dittrich2000zu}. It can be obtained from the imaginary 
part of the three-loop LBL diagrams shown in Fig. \ref{fig-bs}. These diagrams involve the LBL box with two photons on-shell
and two off-shell; the fully off-shell box would appear if one wished to push this calculation to even higher orders. 

\bigskip

\begin{figure}[h]
		\centering 
		\includegraphics[width=.60\textwidth]{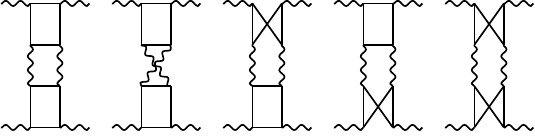}
		\caption{Feynman diagrams contributing to three-loop LBL scattering.}
		\label{fig-wu}
\end{figure}

QED bound-state calculations, too, have reached a level where the LBL box can appear as a subdiagram.  
For example, the Lamb shift gets contributions from the diagrams shown in Fig. \ref{fig-bs} (see \cite{Czarnecki:2016lzl}
and refs. therein). 

\begin{figure}[h]
		\centering 
		\includegraphics[width=.2\textwidth]{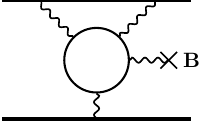}
		\caption{LBL diagram contributing to the bound electron gyromagnetic factor.}
		\label{fig-boundg}
\end{figure}

A similar contribution exists to the bound electron gyromagnetic factor, Fig. \ref{fig-boundg}. It has been computed in \cite{Czarnecki:2020kzi}, albeit only in the Euler-Heisenberg approximation. 

\bigskip

Analogous contributions to the energy levels of positronium are also becoming experimentally relevant \cite{Adkins:2014xoa}.

\subsubsection{Photonic processes involving Coulomb fields}

\begin{figure}[h]
		\centering 
		\includegraphics[scale=1.2]{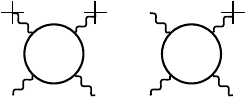}
		\caption{Left-hand diagram: Delbr\"uck scattering. Right-hand diagram: photon splitting.}
		\label{fig-delbruck}
\end{figure}

	There has been indirect experimental evidence for the LBL scattering in processes involving electric fields 
	such as the above-mentioned Delbr\"uck scattering, the elastic scattering of a photon by nuclei that can be realized as a 
	$\gamma\gamma\rightarrow\gamma\gamma$ reaction with two photons from the Coulomb field of nucleus
	(left-hand side of Fig. \ref{fig-delbruck}). Delbr\"uck scattering  
		has been measured for photon energies below 7 GeV in the well-known G\"ottingen experiment \cite{Wilson1953-720,Schumacher1975134,BarNoy1977132}. 
		
		\bigskip
		
The imaginary part of the Delbr\"uck scattering diagram describes the pair-creation process $\gamma + \gamma_{\rm Coulomb} \to e^+ + e^-$ which is presently of great actuality in connection with the so-called ``X17 puzzle'' encountered in beryllium transitions 
(see \cite{Fornal:2017msy,Koch:2020ouk} and refs. therein). 
	
Similarly also photon splitting, the process in which one photon splits into two in the presence  of an external 
field \cite{Shima1966-944} (right-hand side of Fig. \ref{fig-delbruck})
has been observed using Coulomb fields in \cite{Akhmadaliev:2002ik}.
The inverse process of photon merging may be observed in the collision of a high-energy proton beam and a strong laser
field with the next generation of petawatt lasers \cite{DiPiazza:2007prw,DiPiazza:2009cq}. 
	
\subsubsection{Photonic processes involving magnetic fields}
 In the presence of strong external magnetic fields vacuum QED predicts birefringence and dichroism 
due to the modification of the photon dispersion relation 
\cite{Toll:1952rq,Baier:1967zzc,Baier:1967zza,Adler1971599,Dittrich:1998gt,Heyl_1997}
(for a review, see \cite{Battesti:2018bgc}). 
An experiment that focused on the vacuum birefringence effect was PVLAS (Polarization of Vacuum with LASer) \cite{Cameron1993-3707,Bakalov1998-103,Valle2014-092003,Ejlli:2020yhk} but
its measurement has proven elusive. After 25 years, the PVLAS experiment came to an official end in December 2017 (the latest report can be found in \cite{Ejlli:2020yhk}). 
With the emergence of high intensity lasers such as Extreme Light Infrastructure (ELI), Center for Relativistic Laser Science (CoReLS), Helmholtz International Beamline for Extreme Fields (HiBEF) and many more, new proposals involving the utilization of optical and X-ray lasers have surfaced for the study of vacuum birefringence, see  \cite{Heinzl:2006xc,King:2018wtn,Karbstein:2021ldz,Karbstein:2022uwf,Ahmadiniaz:2022nrv} and references therein.

Photon splitting can occur as well in a magnetic field \cite{Adler1971599} which is very important in astrophysics for the study of neutron stars (see \cite{Hu:2019nyw} and references therein) that exhibit a very intense magnetic field. Here, however, the box diagram 
(right-hand side of Fig. \ref{fig-delbruck}) does not contribute in the constant-field approximation because of the peculiar 
collinear kinematics of this process \cite{Adler1971599}. For photon splitting and merging in inhomogeneous fields, see
\cite{Gies:2014jia,Gies:2016czm}.

The fully off-shell LBL box contributes to the anomalous magnetic moment also through diagrams with one leg connecting
to the magnetic field, and in this case already starting from three loops, see the left-hand side of Fig. \ref{fig-g23loop}. 
The diagram shown and the two similar ones arising from the other orderings have been calculated analytically by
S. Laporta and E. Remiddi \cite{Laporta:1993pa}, however through extremely lengthy calculations. 
A major motivation of the present work is that the worldline formalism makes it quite easy to integrate out the low-energy
photon leg connecting to the external field. This ought to lead to substantive simplifications in this type of calculations. 
Similarly, the four-loop contributions to $g-2$ shown on the right-hand side of Fig. \ref{fig-g23loop} have been computed through a heroic effort \cite{Laporta:2017okg,Aoyama:2019ryr}
but the available methods are insensitive to the fact that diagrams that differ only in the ordering of the photons of the light-by-light subdiagram
are related by gauge invariance. 

\subsubsection{Photonic processes involving Coulomb and magnetic fields}

\begin{figure}[h]
		\centering 
		\includegraphics[scale=1.2]{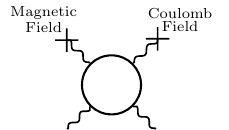}
		\caption{Delbr\"uck scattering for a combined Coulomb and magnetic field.}
		\label{fig-naser}
	\end{figure}

In the recent \cite{Ahmadiniaz:2020kpl} (see also \cite{Ahmadiniaz:2022mcy}), a variant of Delbr\"uck scattering was proposed where a laser beam interacts with both a Coulomb field and a strong external magnetic field, Fig. \ref{fig-naser}. 

The advantage of this configuration is that it yields a large interaction volume, which goes along
with a peak of the differential cross section for the vacuum birefringence signal in the forward direction.

\subsection{Standard model}

While in pure QED usually only the spinor-loop case is considered of direct phenomenological relevance, in QCD
the scalar loop case can appear at low energies with a pion in the loop. In this connection, the
scalar QED four-photon amplitude off-shell 
plays a central role in the dispersive approach to hadronic LBL scattering of \cite{Colangelo:2014dfa}. 
When applied to the hadronic LBL contribution to the muon anomalous magnetic moment, this leads to the left-hand diagram
of Fig. \ref{fig-g23loop} with a pion loop (the same diagram appears with a quark loop in the OPE-based approach to
the hadronic LBL contribution to $g-2$ of \cite{Bijnens:2020xnl,Bijnens:2020yfc,Bijnens:2022qui}).  

The box diagram(s) with two photons and two gluons figure prominently in the recent study of the role of the chiral anomaly
in polarized deeply inelastic scattering by A. Tarasov and R. Venugopalan \cite{Tarasov:2020cwl} (also using the worldline formalism). 
This calculation is still not essentially
different from the purely photonic one because with only two gluons the presence of the color factors does not yet 
make it necessary to fix an ordering of the four legs. 

\begin{figure}[h]
		\centering 
		\includegraphics[width=.30\textwidth]{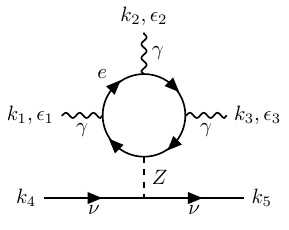}
		\caption{Diagram contributing to the process $\gamma\nu\rightarrow \gamma\gamma\nu$.}
		\label{fig-photonneutrino}
	\end{figure}

Similarly, the computation of the box diagram with three photons and one $Z$ can still be reduced to the one of the
four-photon box since the axial part of the coupling of the $Z$ to the fermion line drops out in the sum of diagrams
\cite{Abada:1998sw,Inan:2021pbx}. 
With the three photons on-shell and the $Z$ off-shell it gives an important contribution to the process
$\gamma\nu\rightarrow \gamma\gamma\nu$ and its crossed processes, see Fig. \ref{fig-photonneutrino}.

\subsection{Beyond the standard model}

The diagram of Fig. \ref{fig-photonneutrino} has also been studied for the left-right symmetric extension of the SM with the $Z$ replaced by a $Z'$
\cite{Abada:1998ek}. 

Another type of exotic particles suggested by BSM models are millicharged fermions. Those would contribute to the 
four-photon amplitude as loop particles in the same way as the standard model fermions. In particular, in a magnetic field they
would lead to vacuum birefringence and dichroism, which has been used in \cite{Gies2006Oct} to obtain strong constraints
on millicharged fermions from existing data on laser experiments.

Another recent application of light-by-light scattering is to the production of dark photons, a new candidate for dark matter \cite{Wong:2021lgk}. 
	
\section{Worldline representation of the $N$-photon amplitudes}\label{WL}

In the worldline approach to scalar QED, the $N$-photon amplitude possesses the following path-integral representation:
\bear
\Gamma_{\rm scal}(k_1,\varepsilon_1;\ldots ; k_N,\varepsilon_N) &=&
(-ie)^N 
\int_0^{\infty}
\frac{dT}{T}\,
\e^{-m^2T}
\int_{x(0)=x(T)}
Dx\,
\e^{-\int_0^T d\tau
\kinb}
\nonumber\\
&& \times
 V_{\rm scal}^{\gamma}[k_1,\varepsilon_1]\cdots V_{\rm scal}^{\gamma}[k_N,\varepsilon_N] \, .
  \nonumber\\
\label{Nphotonvertop}
\ear
Here $T$ is the proper-time of the scalar particle in the loop, and the path integral is performed 
over the space of all closed loops in (Euclidean) spacetime with periodicity $T$.
Each photon is represented by the following {\sl photon vertex operator}, integrated along the trajectory:
\bear
V_{\rm scal}^{\gamma}[k,\varepsilon] &\equiv &\int_0^T d\tau \,\varepsilon\cdot\dot x(\tau)\,{\e}^{ik\cdot x(\tau)}\, .
\label{defphotonvertop}
\ear\no
The path integral is of gaussian form, and thus can be performed by Wick contractions in the 
one-dimensional worldline field theory. Using a formal exponentiation of the factor 
$\varepsilon\cdot\dot x = \e^{\varepsilon\cdot\dot x}\vert_{\varepsilon}$,
one straightforwardly arrives at the following ``Bern-Kosower master formula'':
\begin{eqnarray}
\Gamma_{\rm scal}(k_1,\varepsilon_1;\ldots;k_N,\varepsilon_N)
&=&
{(-ie)}^N
{\int_{0}^{\infty}}\frac{dT}{T}
{(4\pi T)}^{-\frac{D}{2}}
\e^{-m^2T}
\prod_{i=1}^N \int_0^T 
d\tau_i
\nonumber\\
&& \hskip -1.5cm
\times
\exp\biggl\lbrace\sum_{i,j=1}^N 
\Bigl\lbrack  \half G_{ij} k_i\cdot k_j
-i\dot G_{ij}\varepsilon_i\cdot k_j
+\half\ddot G_{ij}\varepsilon_i\cdot\varepsilon_j
\Bigr\rbrack\biggr\rbrace
\Bigl\vert_{\varepsilon_1\varepsilon_2\ldots \varepsilon_N} \, .
\nonumber\\
\label{scalarqedmaster}
\end{eqnarray}
\no
Here the dependence on the proper-time parameters $T,\tau_1,\ldots,\tau_N$ is encoded in the ``worldline Green's function''
\bear
G(\tau_1,\tau_2)=\mid \tau_1-\tau_2\mid 
-\frac{(\tau_1-\tau_2)^2}{T} \,,
\label{1-defGB}
\ear
\no
together with its first and second derivatives:
\begin{eqnarray}
\dot G(\tau_1,\tau_2) &=& {\rm sgn}(\tau_1 - \tau_2)
- 2 \frac{(\tau_1 - \tau_2)}{T} \, ,\nonumber\\
\ddot G(\tau_1,\tau_2)
&=& 2 {\delta}(\tau_1 - \tau_2)
- \frac{2}{T} \, . \quad \nonumber\\
\label{GdGdd}
\end{eqnarray}
\noindent
Dots here denote a
derivative acting on the first variable,
$\dot G(\tau_1,\tau_2) \equiv \frac{\partial}{\partial\tau_1}G(\tau_1,\tau_2)$, 
and we abbreviate
$G_{ij}\equiv G(\tau_i,\tau_j)$ etc.
The notation $\bigl\vert_{\varepsilon_1\varepsilon_2\ldots \varepsilon_N}$ means that, after the expansion of the 
exponential, only those terms should be kept that involve each polarization vector 
$\varepsilon_1,\ldots,\varepsilon_N$ linearly. The fact that we are introducing these polarization vectors does not mean
that we are imposing on-shell conditions; throughout this series of papers we will work fully off-shell, so that all polarization
vectors will be arbitrary four-vectors used for notational convenience only. 

Let us also emphasize already here that \eqref{Nphotonvertop} and \eqref{scalarqedmaster} represent the full $N$-photon amplitude
without the need of summing over permutations of the photon legs. What is more, the permutation (= Bose) symmetry is already manifest
at the integrand level.   

Proceeding to the spinor QED case, differently from the standard Dirac formalism in the worldline formalism the $N$-photon amplitudes
display a manifest break-up into contributions from the orbital and spin degrees of freedom of the loop fermion. The former are
given by the same path integral as in the scalar QED case above, while for the latter there are various equivalent representations
\cite{Feynman:1951gn,Fradkin:1966zz,Gies:2005ke,Edwards:2021vhg}. In the present series of papers, we will follow the approach going back to Fradkin 
\cite{Fradkin:1966zz} 
where \eqref{Nphotonvertop} gets generalized to 
\bear
\Gamma_{\rm spin}(k_1,\varepsilon_1;\ldots ; k_N,\varepsilon_N) &=&
-\half(-ie)^N 
\int_0^{\infty}
\frac{dT}{T}\,
\e^{-m^2T}
\int_P
Dx\,
\e^{-\int_0^T d\tau
\kinb}
\nonumber\\
&& \times
\int_{A} D\psi(\tau)
\,
\e^{
-\int_0^Td\tau
\half \psi\cdot \dot\psi 
}
 V^{\gamma}_{\rm spin}[k_1,\varepsilon_1]\cdots V^{\gamma}_{\rm spin}[k_N,\varepsilon_N]\,,
 \, 
  \nonumber\\
\label{DNpointspin}
\ear
where $P$ stands for ``periodic boundary conditions" and $A$ for ``anti-periodic boundary conditions". 
Now $V^{\gamma}_{\rm spin}$ is the vertex operator for the emission or absorption of a photon by a spin-half particle
(called ``electron'' for definiteness in the following)
\bear
 V^{\gamma}_{\rm spin}[k,\varepsilon] &\equiv &  \int_0^Td\tau\, \Bigl[ \varepsilon\cdot \dot x(\tau)+2i \varepsilon\cdot \psi(\tau) k\cdot\psi(\tau)\Bigr] \,\e^{ik\cdot x(\tau)}
 \, .
\label{defVspin}
\ear
The additional term, representing the interaction of the electron spin with the photon, is written in terms of a
Grassmann-valued Lorentz vector $\psi^\mu (\tau)$ obeying anti-periodic boundary conditions, $\psi^{\mu} (T) = - \psi^{\mu} (0)$.
The Grassmann path integral can be evaluated by Wick contractions using the basic two-point correlator
\bear
\langle \psi^{\mu}(\tau)\psi^{\nu}(\tau')\rangle &=& \half G_F(\tau,\tau')\delta^{\mu\nu} \, ,
\label{wickgrassmann}
\ear
where
\bear
G_F(\tau,\tau') \equiv
{\rm sgn}(\tau - \tau') 
\, .
\label{defGF}
\ear
The free Grassmann path integral is normalized to $4$, the number of real degrees of freedom of the electron/positron 
in four-dimensions.

The worldline representation of the $N$-photon amplitudes has been generalized to include a constant external field in 
\cite{Shaisultanov1996-354,Reuter1997-313}. The generalization to the open-line case in scalar QED, i.e. the scalar propagator dressed with $N$ photons, was given in \cite{daikouji1996bern,Ahmadiniaz:2016-045023}, and extended to the constant-field case in \cite{Ahmad2017-9}. 
Very recently, a computationally efficient worldline representation has also been constructed for the dressed electron propagator \cite{Ahmadiniaz:2020wlm}.
The formalism has also been applied to multiloop QED amplitudes \cite{Schmidt1996-2150,Reuter1997-313,Fliegner:1997ra,Dunne_2002,Dunne:2002qg,Huet:2010nt,Huet:2018ksz,Ahmadiniaz:2016-045023,Nicasio:2020aw}. 

Although in the present series of papers we focus on the QED photon amplitudes, 
it should be mentioned that Bern-Kosower type formulas have been derived also for many other cases. 
See \cite{Bern1991-1669,Bern1992451,Bern1993-2677} for the $N$-gluon amplitudes in QCD,
and \cite{Bastianelli2002-372,Bastianelli2003-104009,AHMADINIAZ2020114877} for amplitudes involving gravitons. 


To start exploiting the master formula, we introduce polynomials $P_N$ 
capturing the result of the expansion of the exponential in \eqref{scalarqedmaster} for fixed $N$:
\bear
\exp\biggl\lbrace 
\cdot
\biggr\rbrace 
\Bigl\vert_{\varepsilon_1\varepsilon_2\ldots \varepsilon_N}
 \quad\equiv
 {(-i)}^N P_N(\dot G_{ij},\ddot G_{ij})
 \exp\biggl[\half \sum_{i,j=1}^N G_{ij}k_i\cdot
k_j \biggr] \, . \label{defPN} \ear\no 
However, in the worldline formalism one does not usually use this form of the integrand. Rather, one applies IBP
to remove all second derivatives $\ddot G_{ij}$ of the worldline Green's function from the integrand. 
This is always possible, and since $G(\tau,\tau')$ is adapted to the periodic boundary conditions does not generate boundary terms.
Thus the whole effect of the IBP is the replacement of $P_N(\dot G_{ij},\ddot G_{ij})$ by a new polynomial 
$Q_N(\dot G_{ij})$. This polynomial is unique for $N=2$ and $N=3$, while beginning with the four-point case different
IBP procedures exist that lead to different equivalent
results \cite{Strassler:1992nc}. This ambiguity was discussed in \cite{Schubert:1997ph} and a definite
algorithm, called ``symmetric partial integration'', given that maintains the full permutation symmetry at each step,
and leads to a unique polynomial $Q_N$. The main advantage from this transition of the ``P-representation'' to the ``Q-representation''
is that the IBP homogenizes the integrand; every term in $Q_N$ has $N$ factors of $\dot G_{ij}$ and $N$ factors of external momentum. This makes it possible to combine terms in the integrand related by gauge invariance, and write it more compactly in terms of
the photon field strength tensors \eqref{fmunu}. Namely, $Q_N$ turns out to contain all possible traces of products of 
field strength tensors, which motivates the definition of a ``Lorentz-cycle'' $Z_n(i_1i_2\ldots i_n)$ as 
\bear
Z_2(ij)&\equiv&
\half {\rm tr}\bigl(f_if_j\bigr) = \varepsilon_i\cdot k_j\varepsilon_j\cdot k_i - \varepsilon_i\cdot\varepsilon_jk_i\cdot k_j \, ,
\nonumber\\
Z_n(i_1i_2\ldots i_n)&\equiv&
{\rm tr}
\Bigl(
\prod_{j=1}^n
f_{i_j}\Bigr)\, , 
\quad (n\geq 3) \, .
\nonumber\\
\label{defZn}
\ear\no
Moreover, it turns out \cite{Strassler:1992nc,Schubert:1997ph}
that in $Q_N$ each Lorentz-cycle $Z_n(i_1i_2\ldots i_n)$ appears multiplied by a corresponding ``$\tau$-cycle''

\bear
\dot G_{i_1i_2} 
\dot G_{i_2i_3} 
\cdots
\dot G_{i_ni_1}
\, .
\label{deftaucycle}
\ear
This motivates the further definition of a ``bicycle'' as the product of the two:

\bear
\dot G(i_1i_2\cdots i_n) \equiv \dot G_{i_1i_2} 
\dot G_{i_2i_3} 
\cdots
\dot G_{i_ni_1}
Z_n(i_1i_2\cdots i_n)
\, .
\label{defbicycle}
\ear
Note that these bicycles are not all independent, since they are invariant under the operations of cyclic permutations and
inversion. Thus  the number of independent cycles of length $n$ is $\frac{n!}{2n}$. For example, for $n=4$ there are only
three really different cycles, which can be chosen as $\dot G(1234)$, $\dot G(1243)$, and $\dot G(1324)$. 

However, $Q_N$ can (except for $N=2$ and in the low-energy limit, see below Section \ref{lowlimit8}) not be written entirely in terms of bicycles and
their products. Starting from $N=3$ there will be leftovers, called ``$n$-tails'' in the notation of \cite{Strassler:1992nc}, where $n$, the 
``length'' of the tail, denotes the number of polarization vectors involved in the tail. In general $Q_N$ will involve tails with length 
$n=1,2,\ldots,N-2$. Thus for our present purposes we will need to know only the one- and two-tails. Those are given by 
	\bear\label{tails}
	T(i)&\equiv&\sum_{r\ne i}\Gd_{ir}\epsk ir\,,\label{1tail} \\
	T(ij)&\equiv&{\sum_{{{r\ne i,s\ne j}}\atop{(r,s)\ne(j,i)}}}
	\Gd_{ir}\epsk ir\Gd_{js}\epsk js+\half\Gd_{ij}\epseps ij\Big\lbrack\sum_{r\ne i,j}\Gd_{ir}\kk ir
	-\sum_{s\ne j,i}\Gd_{js}\kk js\Big\rbrack \,.\label{2tail}\non 
	\ear
Here it should be noted that $\dot G(\tau,\tau)=0$, so that, for example, the term with $r=i$ drops out in the sum defining the one-tail.
In the following, we leave it understood that terms containing a $\Gd_{ii}$ are to be dropped.
For the two-tail, also some terms must be discarded that are non-zero, but would lead to the appearance of a two-cycle in the
tail, and thus to an overcounting. With increasing length of the tails one has to impose an increasing
number of such restrictions to eliminate all the cycles that would otherwise still be present in the tails; 
see \cite{Schubert2001-73} for the details. There also the tails have been worked out explicitly up to length four, which is what is needed for the
six-photon amplitudes
\footnote{Apart from providing compactness, the tails also possess remarkable algebraic properties that become relevant only in the non-abelian case, where they open up
an easy route to the computation of Berends-Giele currents in BCJ gauge \cite{Ahmadiniaz:2021fey}, objects that are central in color-kinematics duality \cite{Bern:2008qj}.}.

After this repackaging of the integrand into cycles and tails, $Q_N$ for sizeable $N$ (which includes already the case $N=4$) becomes
vastly more compact than $P_N$. Another great advantage of the ``Q-representation'' is that it allows one to make the transition
from scalar to spinor QED by a simple pattern-matching procedure, the ``Bern-Kosower replacement rule'' \cite{Bern1991-1669,Bern1992451}:
after the removal of all $\ddot G_{ij}$, the integrand for the $N$-photon amplitude in spinor QED can be obtained from the scalar QED one by multiplying the whole amplitude by a global factor of ``-2" (for statistics and degrees of freedom), and applying the following ``cycle replacement rule'',
	\bear
	\Gd_{i_1i_2}\Gd_{i_2i_3}\cdots\Gd_{i_{n}i_1} \rightarrow \Gd_{i_1i_2}\Gd_{i_2i_3}\cdots\Gd_{i_{n}i_1}- G_{Fi_1i_2}G_{Fi_2i_3}\cdots G_{Fi_{n}i_1} \,,
	\label{rep-rul}
	\ear
	which transforms the bicycle $\Gd(i_1i_2\cdots i_n)$ into the ``super-bicycle"
	\begin{equation}
	\Gd_{\rm S}(i_1i_2\cdots i_n)= \Big(\Gd_{i_1i_2} \Gd_{i_2i_3} \cdots\Gd_{i_ni_1} - G_{Fi_1i_2} G_{Fi_2i_3}\cdots G_{Fi_{n}i_1} \Big) Z_n(i_1 \cdots i_n) \,.
	\label{defsuperbicycle}
	\end{equation}
See \cite{Schubert2001-73} for a proof that this rule correctly generates all the terms that in a direct calculation of the double path integral
\eqref{DNpointspin} would come from Wick contractions using \eqref{wickgrassmann}. The replacement rule has its origin in
a supersymmetry between the orbital and spin degrees of freedom of the electron, called ``worldline supersymmetry'' 
\cite{Strassler:1992nc,Schubert2001-73} which generalizes the well-known supersymmetry of the non-relativistic Pauli-equation
(in the original string-based approach of Bern and Kosower the rule was derived from worldsheet supersymmetry \cite{Bern1991-1669,Bern1992451}).

Alternatively, the worldline supersymmetry can also be used in a more direct way by the introduction of a worldline super formalism
\cite{Strassler1993,Schubert2001-73,polyakov1987gauge}. This allows one to write down the following 
spinor-QED analogue of the master formula \eqref{scalarqedmaster},

\begin{eqnarray}
\Gamma_{\rm spin}
(k_1,\varepsilon_1;\ldots;k_N,\varepsilon_N)
&=&
-2
{(-ie)}^N
\int_{0}^{\infty}{dT\over T}
{(4\pi T)}^{-{D\over 2}}\e^{-m^2T}
\nonumber\\&&
\!\!\!\!\!\!\!\!\!\!\!\!\!\!\!\!\!\!\!\!
\!\!\!\!\!\!\!\!\!\!\!
\!\!\!\!\!\!\!\!\!\!\!\!\!\!\!\!\!\!\!\!
\!\!\!\!\!\!\!\!\!\!\!\!\!\!\!\!\!\!\!\!
\times
\prod_{i=1}^N \int_0^T 
d\tau_i
\int
d\theta_i
\exp\biggl\lbrace
\sum_{i,j=1}^N
\Biggl\lbrack
\half\hat G_{ij} k_i\cdot k_j
+iD_i\hat G_{ij}\varepsilon_i\cdot k_j
+\half D_iD_j\hat G_{ij}\varepsilon_i\cdot\varepsilon_j\Biggr]
\biggr\rbrace
\mid_{\varepsilon_1\ldots\varepsilon_N}\, .
\nonumber\\
\label{supermaster}
\end{eqnarray}
Here the $\theta_i$ are Grassmann variables, 
\bear
\hat G(\tau,\theta;\tau',\theta')
\equiv G_B(\tau,\tau') + \theta\theta' G_F(\tau,\tau')\,,
\label{superpropagator}
\ear\no
is the ``worldline superpropagator'', and
$
D\equiv  {\partial\over{\partial\theta}} - 
   \theta
{\partial\over{\partial\tau}} \label{defD}
$
the super derivative.

\section{The four-photon case}

In the $N=4$ case, a direct expansion of the master formula \eqref{scalarqedmaster} yields the 
``P-representation'' of the scalar QED four-photon amplitude in $D$ dimensions:
\bear
	\Gamma_{\rm scal}(k_1,\veps_1;\ldots ;k_4,\veps_4)
	&=& e^4
	\int_0^\infty \frac{dT}{T}(4\pi T)^{-\frac{D}{2}}\e^{-m^2T}\non
	&&\times\int_0^T\prod_{i=1}^4d\t i\,P_4(\Gd_{ij},\Gdd_{ij})\,\exp\left[\sum_{i,j=1}^4\half G_{ij}\kk ij\right]\,,\non
	\label{4-photonP}
	\ear
	where 
\bear
P_4(\Gd_{ij},\Gdd_{ij})=&&\Gdd_{12}\epseps12 \,\Gdd_{34}\epseps34+\Gdd_{13}\epseps13\,\Gdd_{24}\epseps24+\Gdd_{14}\epseps14\,\Gdd_{23}\epseps23\non
&-&\Gd_{1i}\epsk1i\,\Gd_{2j}\epsk2j\,\Gdd_{34}\epseps34-\Gd_{1i}\epsk1i\,\Gd_{3j}\epsk3j\,\Gdd_{24}\epseps24\non
&-&\Gd_{1i}\epsk1i\,\Gd_{4j}\epsk4j\,\Gdd_{23}\epseps23-\Gd_{2i}\epsk2i\,\Gd_{3j}\epsk3j\,\Gdd_{14}\epseps14\non
&-&\Gd_{2i}\epsk2i\,\Gd_{4j}\epsk4j\,\Gdd_{13}\epseps13-\Gd_{3i}\epsk3i\,\Gd_{4j}\epsk4j\,\Gdd_{12}\epseps12\non
&+&\Gd_{1i}\epsk1i\,\Gd_{2j}\epsk2j\,\Gd_{3k}\epsk3k\,\Gd_{4l}\epsk4l\,.\non
\label{P_4}
\ear
and the indices $i,j,k,l$ are to be summed from $1$ to $4$. 
As described above, by IBP this can be transformed into the corresponding ``Q-representation'',
	\bear
	\Gamma_{\rm scal}(k_1,\veps_1;\ldots ;k_4,\veps_4)
	&=& e^4
	\int_0^\infty \frac{dT}{T}(4\pi T)^{-\frac{D}{2}}\e^{-m^2T}
	\int_0^T\prod_{i=1}^4d\t i\,Q_4(\Gd_{ij})\,\exp\left[\sum_{i,j=1}^4\half G_{ij}\kk ij\right]\,,\non
	\label{4-photonQ}
	\ear
	where now 
	\footnote{When comparing with \cite{Schubert2001-73} note that there a different basis was used for the four-cycle component $Q^4_4$. The two bases are related by cyclicity and inversion.}
	\bear\label{Q_4}
	Q_4&=&Q^4_4+Q^3_4+Q^2_4+Q^{22}_{4}\,,\non
	Q^4_4&=&\Gd(1234)+\Gd(2314)+\Gd(3124)\,,\non
	Q^3_4&=&\Gd(123)T(4)+\Gd(234)T(1)+\Gd(341)T(2)+\Gd(412)T(3)\,,\non
	Q^2_4&=&\Gd(12)T(34)+\Gd(13)T(24)+\Gd(14)T(23)+\Gd(23)T(14)+\Gd(24)T(13)+\Gd(34)T(12)\,,\non
	Q^{22}_4&=&\Gd(12)\Gd(34)+\Gd(13)\Gd(24)+\Gd(14)\Gd(23)\, .\non
	\label{q4}
\ear
Although this is already an extremely nice and compact representation of this amplitude, at this stage it is natural to ask whether
further IBP might exist that would allow one to absorb into field-strength tensors not only the
polarization vectors contained in the cycles, but also those in the tails. It turns out that this is indeed possible, 
at the cost of introducing inverse factors of momenta in the IBP, and a set of (almost) arbitrary auxiliary vectors $r_i^{\mu}$. 
In \cite{Ahmadiniaz2013-132} it was shown how to rewrite, by suitable such IBPs involving only the tail variables, 
a tail of arbitrary length completely in terms of
field strength tensors. The resulting representation of the scalar and spinor QED photon amplitudes, which has the remarkable property of being manifestly transversal already at the integrand level, was called ``QR-representation''.
For our purposes, we will need only the corresponding form of the one-tail:
\bear
T_R(i) &\equiv& \sum_{j\ne i} \Gd_{ij}~ \frac{r_i \cdot f_i \cdot k_j}{r_i \cdot k_i} \, .
\label{1tailr}
\ear
The vector $r_i$, called ``reference vector'', is arbitrary except that it has to obey $r_i\cdot k_i\ne 0$.
To obtain \eqref{1tailr} from \eqref{1tail}, one needs a single total derivative term:
\bear
T_R(i) \,\e^{(\cdot)} = T(i)\edot - \partial_i \Bigl( \frac{r_i\cdot\varepsilon_i}{r_i\cdot k_i} \,\e^{(\cdot)} \Bigr)\,.
\ear
Here we have introduced the abbreviations $\partial_i \equiv \frac{d}{d\tau_i}$ and
	\bear
	\e^{(\cdot)}\equiv \exp\left[\sum_{i,j=1}^4\half G_{ij}\kk ij\right]\,.
	\ear
The choice of the reference vector for most purposes should be done in such a way that the permutation symmetry between
the photons with index other than $i$ remains unbroken. Since one usually would like to avoid introducing new vectors,
a natural choice is $r_i = k_i$, however this creates a $k_i^2$ in the denominator, and thus a spurious singularity in the
on-shell limit. Alternatively, one could use each of the remaining photon momenta and take the average of the
resulting integrals. 

For the two-tail, various versions were obtained in \cite{Ahmadiniaz2013-132}, but here we will rather introduce a new one, 
the ``short two-tail'' $T_{sh}(ij)$, that we have found much superior to those. It is defined by
\bear
T_{sh}(ij) \equiv 
\sum_{r,s\ne i,j}
	\Gd_{ri} \Gd_{js}~ \frac{k_r \cdot f_i\cdot f_j\cdot k_s}{k_i\cdot k_j}\,,
\label{2tails}
\ear
and obtained from the Q-version of the two-tail \eqref{2tail} by adding the following set of total derivative terms,
	\bear
T_{sh}(ij)\edot &=& T(ij)\edot +\frac{1}{k_i\cdot k_j}
\biggl\lbrack
\varepsilon_i\cdot\varepsilon_j\,\partial_i\partial_j\edot
- \varepsilon_i\cdot k_j\varepsilon_j\cdot k_s\partial_i\Bigl(\dot G_{js}\edot\Bigr)
\nonu &&
- \varepsilon_j\cdot k_i\varepsilon_i\cdot k_r\partial_j\Bigl(\dot G_{ir}\edot\Bigr)
 + \Bigl(\half\varepsilon_i \cdot \varepsilon_j k_i\cdot k_j - \varepsilon_i\cdot k_j \varepsilon_j\cdot k_i\Bigr) 
(\partial_i-\partial_j) \Bigl( \dot G_{ij}  \,\e^{(\cdot)}\Bigr )
\biggr\rbrack \,,
\nonumber\\
	\ear
($i,j$ fixed, $r,s$ summed over).
Note that the indices $r,s$ in \eqref{2tails} now take only values in the cycle variables, thus there are only
four terms instead of twelve as compared to the Q-tail \eqref{2tail}. This explains the name ``short tail''. 
Thus when we have the short two-tail and the Q one-tail the representation will be called ``STQ-representation''. The manifestly transversal representation comes with the short two-tail and the R one-tail, being then the ``STR-representation''. 
Let us process the integrand a bit further by the standard rescaling
	 \bear
	 \tau_i = Tu_i, \quad  i=1,\ldots,4 \, .
	 \label{rescale}
	 \ear
	 We can then write the four-photon amplitudes, combined for scalar and spinor QED, in the following way:
	 \bear
	\Gamma_{\{{{\rm scal}\atop{\rm spin}}\}}(k_1,\veps_1;\ldots ;k_4,\veps_4)
	&=& 
	\Big\{{{1}\atop {-2}}\Big\}
	\frac{e^4}{(4\pi)^{\frac{D}{2}}}
	\int_0^\infty \frac{dT}{T} T^{4-\frac{D}{2}}\e^{-m^2T}
	\int_0^1\prod_{i=1}^4du_i\,Q_{\{{{\rm scal}\atop{\rm spin}}\}}(\Gd_{ij})\,
	\e^{(\cdot)} \, .
	\nonumber\\
	\label{4-photon}
	\ear
	The prefactor polynomial for scalar QED, $Q_{\rm scal}(\Gd_{ij})$, is given by \eqref{q4} with the 
	cycles defined in \eqref{defbicycle} and the tails defined in \eqref{1tailr} and \eqref{2tails}
	(here and in the following we will often drop the subscript `4' on $Q$). The one for spinor QED, $Q_{\rm spin}(\Gd_{ij})$,
	has all bicycle factors replaced by the super-bicycle factors \eqref{defsuperbicycle}. 
		Note that we do not introduce a new symbol for the worldline Green's function after the rescaling, which is now given
	by
	\bear
	G_{ij} = |u_i-u_j| - (u_i-u_j)^2 \,.
	\label{defGrescaled}
	\ear
	Note further that $T$ appears linearly in the rescaled exponent,
	\bear
	\e^{(\cdot)} \equiv  \exp\left[\sum_{i,j=1}^4 \frac{T}{2} G_{ij}\kk ij\right]\,.
	\ear
	It should be clear from the context whether the original or the rescaled Green's function is to be used. 
		We will also often strip off the global prefactors of the amplitude, defining
	\bear
	\Gamma_{\{{{\rm scal}\atop{\rm spin}}\}}
	&\equiv & 
	\Big\{{{1}\atop {-2}}\Big\}
	\frac{e^4}{(4\pi)^{\frac{D}{2}}}
	\hat\Gamma_{\{{{\rm scal}\atop{\rm spin}}\}}\,.
\label{defhatGamma}
\ear
	
\section{Worldline vs. Feynman diagram representations}

Next, let us discuss the relation between the parameter integrals appearing
in the worldline representation of the $N$-photon amplitudes and the
parameter integrals arising in the calculation of the corresponding Feynman diagrams
using Feynman-Schwinger parameters.  

In scalar QED, the parameter integrals of the P-representation, obtained directly from the expansion of the
master formula \eqref{scalarqedmaster}, can, for a fixed ordering of the legs, 
straightforwardly be identified with the ones of the Feynman-Schwinger parameter integral
representation of the corresponding Feynman diagrams. 
\begin{figure}[h]
\centering
    \includegraphics[scale=1]{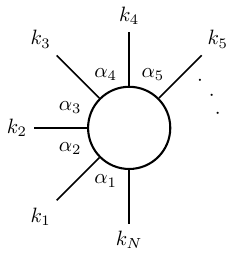}
    \caption      {Transformation to Feynman-Schwinger parameters.}
    \label{fig-Npoint}
  \end{figure}
E.g. for the ``standard'' ordering of the legs the transformation can be read off from Fig. \ref{fig-Npoint}.
Using the translation invariance in the proper-time to set $\tau_N=0$, and choosing the parametrization
of the loop such that $T\geq\tau_1\geq\tau_2\geq\ldots\tau_N=0$, one gets

\bear
\tau_1 &=& \alpha_2 + \alpha_3 + \ldots + \alpha_N \,,\nonu
\tau_2 &=& \alpha_3 + \alpha_4 + \ldots + \alpha_N \,,\nonu
\vdots \, & & \quad \vdots \nonu
\tau_{N-1} &=& \alpha_N \,.\nonu
\label{tautoalpha}
\ear
The $\delta(\tau_i-\tau_j)$ in a $\ddot G_{ij}$ brings together photons $i$ and $j$ and thus creates a quartic vertex
that corresponds to the seagull vertex in the standard formalism. It contributes only if legs $i$ and $j$ are adjacent 
for this ordering, and has to be split equally between the two sectors $\tau_i > \tau_j$ and $\tau_i < \tau_j$. 
%

In our $N=4$ case, the transformation \eqref{tautoalpha}
leads to the following formulas for the various worldline Green's functions and their derivatives: 

\bear
\G12&=&\alpha_2(1-\alpha_2)\hspace{2.2cm},\hspace{0.5cm}\Gd_{12}=1-2\alpha_2\,,\non
\G13&=&(\alpha_1+\alpha_4)(1-\alpha_1-\alpha_4)~~~,\hspace{0.3cm}\Gd_{31}=1-2(\alpha_1+\alpha_4)\,,\non
\G14&=&\alpha_1(1-\alpha_1)\hspace{2.2cm},\hspace{0.5cm}\Gd_{41}=1-2\alpha_1\,,\non
\G23&=&\alpha_3(1-\alpha_3)\hspace{2.2cm},\hspace{0.5cm}\Gd_{23}=1-2\alpha_3\,,\non
\G24&=&(\alpha_1+\alpha_2)(1-\alpha_1-\alpha_2)\hspace{0.3cm},\hspace{0.3cm}\Gd_{24}=1-2(\alpha_3+\alpha_4)\,,\non
\G34&=&\alpha_4(1-\alpha_4)\hspace{2.2cm},\hspace{0.5cm}\Gd_{34}=1-2\alpha_4\,,\non
\ear
and the exponential factor can, using momentum conservation, be identified with the standard four-point denominator of the Feynman-Schwinger
representation,
\bear
\sum_{i<j=1}^4 G_{ij}\kk ij&=&-\alpha_1\alpha_2k_1^2-\alpha_2\alpha_3k_2^2-\alpha_3\alpha_4k_3^2-\alpha_1\alpha_4k_4^2-\alpha_2\alpha_4(k_1+k_4)^2-\alpha_1\alpha_3(k_1+k_2)^2 \,. \nonu
\label{exponent}
\ear
Setting $\tau_4=0$, splitting the remaining three $\tau$-integrals into the six ordered sectors

\begin{figure}[htbp]
\begin{center}
\includegraphics[scale=0.85]{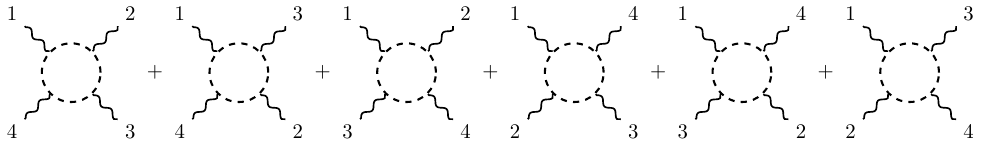}
\caption{Six permuted Feynman diagrams for photon-photon scattering.}
\label{fig-photonphoton}
\end{center}
\end{figure}

\begin{figure}[htbp]
\begin{center}
\includegraphics[scale=0.95]{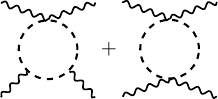}
\caption{The two types of photon-photon scattering diagrams in scalar QED involving the seagull vertex.}
\label{fig-photonphotonseagull}
\end{center}
\end{figure}

\bear
{\rm sector} \,1&:&  \tau_1\geq\tau_2\geq\tau_3\geq\tau_4=0 \,, \nonu
{\rm sector}\, 2&:&  \tau_1\geq\tau_3\geq\tau_2\geq\tau_4=0 \,, \nonu
{\rm sector}\, 3&:&  \tau_3\geq\tau_1\geq\tau_2\geq\tau_4=0 \,, \nonu
{\rm sector}\, 4&:&  \tau_3\geq\tau_2\geq\tau_1\geq\tau_4=0 \,, \nonu
{\rm sector}\, 5&:&  \tau_2\geq\tau_3\geq\tau_1\geq\tau_4=0 \,, \nonu
{\rm sector}\, 6&:&  \tau_2\geq\tau_1\geq\tau_3\geq\tau_4=0 \,, \nonu
\label{defsectors}
\ear
and applying in each sector the appropriate transformation rule, one would arrive at the same collection of
$\alpha$ - parameter integrals as in the calculation of  the familiar sum of six diagrams shown in Fig. \ref{fig-photonphoton}
\noindent
and the terms involving $\delta$ functions would correspond precisely to the additional diagrams that one has in the scalar QED
case involving one or two seagull vertices, see Fig. \ref{fig-photonphotonseagull}. 

This makes it also clear that, in a straightforward use of the P-representation along these lines, we would not achieve much of
an advantage over a standard Feynman-Schwinger parameter calculation. 
To improve on the standard method, it is essential to recognize,
following \cite{Bern1991-1669,Bern1992451,Strassler:1992nc}, that the integrand of the P-representation is written in a form that suggests
a removal of the $\ddot G_{ij}$s by IBP. The resulting Q-representation \eqref{4-photonQ} possesses already 
very substantive advantages over the standard one, which we can summarize as follows:

\benn

\item
It represent already the complete four-photon amplitude in scalar QED, corresponding to the sum of all the Feynman diagrams
shown in Figs. \ref{fig-photonphoton} and \ref{fig-photonphotonseagull}. 

\item
When rewritten in terms of Feynman-Schwinger parameters, the worldline integrals will still be of the 
same type that one would encounter in a Feynman diagram calculation. 
Nevertheless, due to the rearrangement into cycles and tails
the worldline integrand will still be much more compact than what one would find in any straightforward
diagrammatic calculation. 

\item
As is well-known, the four-photon amplitudes in both scalar and spinor QED suffer from spurious UV divergences that cancel out in the sum of all Feynman diagrams by gauge invariance 
\footnote{It is well-known that in QED gauge invariance can get lost if divergences are not handled with due care. See, e.g., \cite{Liang:2012sj}.}. 
In the P-representation, those divergences are still there and contained in the terms 
involving products of two $\ddot G_{ij}$s, but they get removed by the IBP procedure; in the Q-representation all integrals are already
finite term by term. 

\item
The arrangement into cycles and tails allows one to obtain the corresponding spinor QED
amplitude trivially by the application of the replacement rule (\ref{rep-rul}). Thus effectively one bypasses
all the Dirac-algebra manipulations of the standard approach.

\enn

After transforming to the improved tails \eqref{1tailr} and \eqref{2tails}, 
one gets the further advantage of manifest gauge invariance at the integrand level. 

\section{Minimal tensor bases for the four-photon amplitudes}

For off-shell amplitudes, it is usually useful to have a tensor decomposition which is minimal (without redundancies) and
compatible with all their symmetries. For our four-photon amplitude, the only symmetry around is the permutation (= Bose) 
symmetry between the photons. Thus we will now study how many independent tensor structures are contained in the
worldline representation of the amplitude. 

Clearly, in the decomposition \eqref{q4} the terms $Q^4$, $Q^3$ and $Q^{22}$ count only as one structure each, once permutations are taken into account. Thus only $Q^2$ needs to be analyzed, and here all six terms are related by permutations, so it is sufficient to
analyze the first one, that is $\Gd(12)T_{sh}(34)$ in the scalar and  $\Gd_S(12)T_{sh}(34)$ in the spinor QED case. Writing out
$T_{sh}(34)$ 

\bear
T_{sh}(34)&=&
\sum_{r,s=1,2} \Gd_{r3} \Gd_{4s}~ \frac{k_r \cdot f_3 \cdot f_{4} \cdot k_s}{k_3\cdot k_4}\,,
\ear
gives four terms, but they are not all independent, 
due to the symmetry $1\leftrightarrow 2$ of the cycle factor $\Gd_{(S)}(12)$
that leads to the equivalences

\bear
r=s=1 &\sim& r=s=2 \label{equivalence1} \,, \nonumber\\
r=1,s=2 &\sim& r=2,s=1 \label{equivalence2} \,. \nonumber\\
\ear
Thus the total number of independent tensor structures is five.
We will label them in the following way:
\bear
T^{(1)}_{(1234)} & \equiv & Z_4(1234) \, ,  \nonumber\\
T^{(2)}_{(12)(34)} & \equiv & Z_2(12)Z_2(34) \, , \nonumber\\
T^{(3)r_4}_{(123)i} & \equiv & Z_3(123) \frac{r_4\cdot f_4\cdot k_i }{r_4\cdot k_4}\, , \quad i=1,2,3 \, , \nonumber\\
T^{(4)}_{(12)ii} &\equiv & Z_2(12) \,\frac{k_i \cdot f_3 \cdot f_{4} \cdot k_i}{k_3\cdot k_4}\, , \quad i=1,2 , \nonumber\\
T^{(5)}_{(12)ij} &\equiv & Z_2(12) \, \frac{k_i \cdot f_3 \cdot f_{4} \cdot k_j}{k_3\cdot k_4}\,  ,\quad (i,j)=(1,2),(2,1)  .\nonumber\\
\label{defTi}
\ear  
For completeness, let us also write down the full set of permutations of these tensors that will appear in the tensor decomposition:
\bear
&&T^{(1)}_{(1234)},T^{(1)}_{(1243)},T^{(1)}_{(1324)}, 
T^{(2)}_{(12)(34)}, T^{(2)}_{(13)(24)}, T^{(2)}_{(14)(23)},
\nonumber\\
&& T^{(3)r_4}_{(123)1}, T^{(3)r_4}_{(123)2},T^{(3)r_4}_{(123)3},
T^{(3)r_1}_{(234)2}, T^{(3)r_1}_{(234)3},T^{(3)r_1}_{(234)4},
T^{(3)r_2}_{(341)3}, T^{(3)r_2}_{(341)4},T^{(3)r_2}_{(341)1},
T^{(3)r_3}_{(4,1,2)4}, T^{(3)r_3}_{(4,1,2)1}, T^{(3)r_3}_{(4,1,2)2}, 
\nonumber\\
&&T^{(4)}_{(12)11}, T^{(4)}_{(12)22}, T^{(4)}_{(13)11}, T^{(4)}_{(13)33}, T^{(4)}_{(14)11}, T^{(4)}_{(14)44},
T^{(4)}_{(23)22}, T^{(4)}_{(23)33}, T^{(4)}_{(24)22}, T^{(4)}_{(24)44}, T^{(4)}_{(34)33}, T^{(4)}_{(34)44}, \nonumber\\
&&T^{(5)}_{(12)12}, T^{(5)}_{(12)21}, T^{(5)}_{(13)13}, T^{(5)}_{(13)31}, T^{(5)}_{(14)14}, T^{(5)}_{(14)41},
T^{(5)}_{(23)23}, T^{(5)}_{(23)32}, T^{(5)}_{(24)24}, T^{(5)}_{(24)42}, T^{(5)}_{(34)34}, T^{(5)}_{(34)43}\, . \nonumber\\
\ear         
Note that these five tensors (we call them tensors since eventually the polarization vectors have to be stripped off)
all have the same mass dimension, and that $T^{(3)r_4}_{(123)i}$ still depends on the choice of a reference momentum $r_4$ 
(which must {\it not} depend on $i$). 
$T^{(1)}$ and $T^{(2)}$ appear in three permutations, $T^{(3)}$, $T^{(4)}$ and $T^{(5)}$ in twelve each. 
The complete four-photon amplitude can thus be decomposed as 
\bear
\hat \Gamma &=& \hat \Gamma^{(1)} + \hat \Gamma^{(2)}  + \hat \Gamma^{(3)}  + \hat \Gamma^{(4)}  + \hat \Gamma^{(5)} \,,
\label{Gammadecomp}
\ear
\bear
\hat \Gamma^{(1)} &=& \hat \Gamma^{(1)}_{(1234)}T^{(1)}_{(1234)}
 + 
  \hat \Gamma^{(1)}_{(1243)}T^{(1)}_{(1243)}
   + 
  \hat \Gamma^{(1)}_{(1324)}T^{(1)}_{(1324)} \,,
  \nonu
  \hat \Gamma^{(2)} &=& \hat \Gamma^{(2)}_{(12)(34)}T^{(2)}_{(12)(34)}
 + 
  \hat \Gamma^{(2)}_{(13)(24)}T^{(2)}_{(13)(24)}
   + 
  \hat \Gamma^{(2)}_{(14)(23)}T^{(2)}_{(14)(23)} \,,
  \nonu
  \hat \Gamma^{(3)} &=&  \sum_{i=1,2,3} \hat \Gamma^{(3)}_{(123)i}T^{(3)r_4}_{(123)i}
  +
  \sum_{i=2,3,4} \hat \Gamma^{(3)}_{(234)i}T^{(3)r_1}_{(234)i}
  +
  \sum_{i=3,4,1} \hat \Gamma^{(3)}_{(341)i}T^{(3)r_2}_{(341)i}
  +
 \sum_{i=4,1,2} \hat \Gamma^{(3)}_{(412)i}T^{(3)r_3}_{(412)i} \,,
  \nonu
  \hat \Gamma^{(4)} &=&  
\sum_{i<j}  \hat \Gamma^{(4)}_{(ij)ii}T^{(4)}_{(ij)ii} +
\sum_{i<j}  \hat \Gamma^{(4)}_{(ij)jj}T^{(4)}_{(ij)jj} \,,
\nonu
  \hat \Gamma^{(5)} &=&  
\sum_{i<j}  \hat \Gamma^{(5)}_{(ij)ij}T^{(5)}_{(ij)ij} +
\sum_{i<j}  \hat \Gamma^{(5)}_{(ij)ji}T^{(5)}_{(ij)ji} \,.
\nonu
\label{Gammadecompparts}
\ear
The coefficient functions are given, with the convention \eqref{defhatGamma}, by 
\bear
	\hat\Gamma^{(k)}_{\cdots}
	&=& 
	\int_0^\infty \frac{dT}{T} T^{4-\frac{D}{2}}\e^{-m^2T}
	\int_0^1\prod_{i=1}^4du_i\, \hat \gamma^{(k)}_{\ldots}(\Gd_{ij})\,
	\e^{(\cdot)}\,,
	\nonumber\\
	\label{gamma}
	\ear
where, for spinor QED,
\bear
\hat \gamma^{(1)}_{(1234)} &=& \Gd_{12}\Gd_{23}\Gd_{34}\Gd_{41} - G_{F12}G_{F23}G_{F34}G_{F41} \,, \nonu
\hat \gamma^{(2)}_{(12)(34)} &=& \bigl(\Gd_{12}\Gd_{21} - G_{F12}G_{F21}\bigr)  \bigl(\Gd_{34}\Gd_{43} - G_{F34}G_{F43}\bigr) \,, \nonu
\hat \gamma^{(3)}_{(123)1} &=& \bigl(\Gd_{12}\Gd_{23}\Gd_{31} - G_{F12}G_{F23}G_{F31}\bigr) \Gd_{41} \,, \nonu
\hat \gamma^{(4)}_{(12)11} &=&  \bigl(\Gd_{12}\Gd_{21} - G_{F12}G_{F21}\bigr)  \Gd_{13}\Gd_{41} \,, \nonu
\hat \gamma^{(5)}_{(12)12} &=&  \bigl(\Gd_{12}\Gd_{21} - G_{F12}G_{F21}\bigr)  \Gd_{13}\Gd_{42} \nonu
\label{hatgamma}
\ear
(plus permutations thereof), 
and the coefficient functions for scalar QED are obtained from these simply by deleting all the $G_{Fij}$.

A further option is to use the trivial but highly useful identity
\bear
\dot G_{ij}^2 = 1 - 4 G_{ij}\,,
\label{idtriv}
\ear
to rewrite
\bear
\Gd_{ij}\Gd_{ji} - G_{Fij}G_{Fji} 
=
4 G_{ij}\,.
\label{rewrite}
\ear
This will be useful not only for compactness, but also because factors of $G_{ij}$ can be generated by derivatives
applied to the universal exponential factor. 
We can then replace \eqref{hatgamma} by the remarkably compact expressions
\bear
\hat \gamma^{(1)}_{(1234)} &=& \Gd_{12}\Gd_{23}\Gd_{34}\Gd_{41} - G_{F12}G_{F23}G_{F34}G_{F41} \,, \nonu
\hat \gamma^{(2)}_{(12)(34)} &=& 16G_{12}G_{34} \,, \nonu
\hat \gamma^{(3)}_{(123)1} &=& \bigl(\Gd_{12}\Gd_{23}\Gd_{31} - G_{F12}G_{F23}G_{F31}\bigr) \Gd_{41} \,, \nonu
\hat \gamma^{(4)}_{(12)11} &=&  4G_{12}  \Gd_{13}\Gd_{41} \,, \nonu
\hat \gamma^{(5)}_{(12)12} &=& 4G_{12} \Gd_{13}\Gd_{42} \,. \nonu
\label{hatgammanew}
\ear

Finally, let us compare our basis of invariants with the one obtained in \cite{Costantini:1971cj} in a very different way,
namely by writing down all possible tensor structures and systematically reducing them by the application of the
Ward identity and the permutation group. This also led them to a basis composed of five
gauge-invariant tensors $I_{\mu \nu \rho \sigma}^{(i)}$, chosen as 

	\begin{equation}
	I_{\mu \nu \rho \sigma}^{(1)}(1234) = \frac{1}{32} F_{\mu \alpha \beta}(1) F_{\nu \beta \alpha}(2) F_{\rho \gamma \delta}(3) F_{\sigma \delta \gamma}(4) \,,
	\end{equation}
	\begin{equation}
	I_{\mu \nu \rho \sigma}^{(2)}(1234) = \frac{1}{8} F_{\mu \alpha \beta}(1) F_{\nu \beta \gamma}(2) F_{\rho \gamma \delta}(3) F_{\sigma \delta \alpha}(4) \,,
	\end{equation}
	\begin{equation}
	I_{\mu \nu \rho \sigma}^{(3)}(1234) = -\frac{1}{4(k^{(3)} k^{(4)})} F_{\mu \alpha \beta}(1) F_{\nu \beta \alpha}(2) k_{\gamma}^{} F_{\rho \gamma \delta}(3) F_{\sigma \delta \xi}(4) k_{\xi}^{(1)} \,,
	\end{equation}
	\begin{equation}
	I_{\mu \nu \rho \sigma}^{(4)}(1234) = -\frac{1}{4(k^{(3)} k^{(4)})} F_{\mu \alpha \beta}(1) F_{\nu \beta \alpha}(2) k_{\gamma}^{(2)} F_{\rho \gamma \delta}(3) F_{\sigma \delta \xi}(4) k_{\xi}^{(1)} \,,
	\end{equation}
	\begin{equation}
	I_{\mu \nu \rho \sigma}^{(5)}(1234) = \frac{1}{3(k^{(2)} k^{(4)})} k_{\alpha}^{(2)} F_{\sigma \alpha \beta}(4) F_{\mu \beta \gamma}(1) \Big[ F_{\nu \gamma \delta}(2) F_{\rho \delta \xi}(3) - F_{\rho \gamma \delta}(3) F_{\nu \delta \xi}(2) \Big] k_{\xi}^{(1)}\,.
	\end{equation}
	Here the third-rank tensor $F_{\eta \mu \nu}(i)$ is defined by
	\begin{equation}
	F_{\eta \mu \nu}(i) = - F_{\eta \nu \mu}(i) = \delta_{\eta \mu} k_{\nu}^{(i)} - \delta_{\eta \nu} k_{\mu}^{(i)}\,,
	\end{equation}
	which is related to (\ref{fmunu}) by
	\begin{equation}
	\varepsilon_\eta^{(i)} F_{\eta \mu \nu}(i) = f_i^{\nu \mu} \,.
	\end{equation}
	Let us define	 
	\begin{equation}
	I^{(i)} \equiv \varepsilon_\mu^{(1)} \varepsilon_\nu^{(2)} \varepsilon_\rho^{(3)} \varepsilon_\sigma^{(4)} I_{\mu \nu \rho \sigma}^{(i)} \,.
	\end{equation}
	We can 
	then write the five gauge-invariant tensors $I^{(i)}$ in terms of the field strength tensors $f_i$ as follows:
	\begin{equation}\label{Iin}
	I^{(1)}(1234) = \frac{1}{32} {\rm tr}(f_1 f_2 ) {\rm tr}(f_3 f_4 ) \,,
	\end{equation}
	\begin{equation}
	I^{(2)}(1234) = \frac{1}{8} {\rm tr}(f_1 f_2 f_3  f_4 ) \,,
	\end{equation}
	\begin{equation}
	I^{(3)}(1234) = -\frac{1}{4 k_{3} \cdot k_{4}} {\rm tr}(f_1 f_2) k_1\cdot f_3\cdot f_4\cdot k_1 \,,
	\end{equation}
	\begin{equation}
	I^{(4)}(1234) = -\frac{1}{4 k_{3} \cdot k_{4}} {\rm tr}(f_1 f_2) k_2\cdot f_3\cdot f_4\cdot k_1 \,,
	\end{equation}
	\begin{equation}\label{Ifin}
	I^{(5)}(1234) = \frac{1}{3 k_{2} \cdot k_{4}} k_2\cdot f_4\cdot k_1 ~{\rm tr}(f_1 f_2 f_3) \,.
	\end{equation}
	
It thus becomes apparent that this basis $\lbrace I^{(i)}\rbrace$ differs from our basis $\lbrace T^{(i)}\rbrace$ 
only by some rescaling:
\bear
T^{(1)}_{(1234)} & = & 8 I^{(2)}(1234) \,, \nonumber\\
T^{(2)}_{(12)(34)} & = & 8  I^{(1)}(1234) \,, \nonumber\\
T^{(3)k_2}_{(123)1} & = & 3 I^{(5)}(1234) \,, \nonumber\\
T^{(4)}_{(12)11} &= & - 2  I^{(3)}(1234) \,, \nonumber\\
T^{(5)}_{(12)12} &= & - 2  I^{(4)}(2134) \,. \nonumber\\
\label{trafoIT}
\ear 
Note, however, that for $T^{(3)}$ to match with $I^{(5)}$ we had to choose the reference momentum of the one-tail as one of the
momenta of the other photons, which is, of course, not obligatory. In this sense our basis is slightly more general
than the one of \cite{Costantini:1971cj}.

We find it remarkable that our IBP procedure, whose main purpose was to write the integrand of the four-photon
amplitudes in the most compact possible way, has led to the emergence of the same tensor basis that was found by 
\cite{Costantini:1971cj} using the Ward identity. 

It must be noted, though, that the price to pay for this optimization is the introduction of the tensors $T^{(4)}$ and $T^{(5)}$
that carry denominators, and thus introduce spurious kinematic singularities. It is easy to see that this cannot be avoided if one
wishes the basis to be minimal as well as manifestly transversal. 
For some purposes it may be preferable to use a tensor basis that is less optimized but 
free of kinematic singularities \cite{Colangelo:2014dfa}.

\section{Matching identities}

Useful identities can be derived by comparing the worldline integrands before and after the IBP procedure. 
The key observation is that the terms involving purely products of $\ddot G_{ij}$ resp. $D_iD_j\hat G_{ij}$
in the master formulas \eqref{scalarqedmaster} and \eqref{supermaster} involve only scalar integrals, and
are thus trivial from the point of view of tensor reduction. We will now work this out for the four-photon case,
using the notation

\bear
\hat I_n^D(k_1,\ldots,k_n) 
&\equiv& \int_0^\infty \frac{dT}{T} T^{n-\frac{D}{2}}\e^{-m^2T}\prod_{i=1}^n \int_0^1 du_i 
\e^{T\sum_{i<j=1}^n G_{ij}k_{ij}} \,,
\nonu
\hat I_n^D(k_1,\ldots,k_n) [X]
&\equiv& \int_0^\infty \frac{dT}{T} T^{n-\frac{D}{2}}\e^{-m^2T}\prod_{i=1}^n \int_0^1 du_i \, X
\e^{T\sum_{i<j=1}^n G_{ij}k_{ij}} \,,
\nonu
\ear 
where we now also abbreviate $k_{ij} \equiv k_i\cdot k_j$ and 
$\varepsilon_{ij} \equiv \varepsilon_i\cdot\varepsilon_j$.

\subsection{Scalar QED}

Thus let us focus on the term involving $\ddot G_{12}\ddot G_{34}$ in the master formula \eqref{scalarqedmaster}
for $N=4$. In the above notation, it can be written in terms of scalar integrals as 
\bear
\varepsilon_{12}\varepsilon_{34} \hat  I_4^D(k_1,\ldots,k_4) [\ddot G_{12}\ddot G_{34}]
&=&
\varepsilon_{12}\varepsilon_{34} 
\hat I_{\rm scal} \,,
\label{GddGdd}
\ear
where
\bear
\hat I_{\rm scal} &=& 4 \Bigl[\hat I_2^D(k_1+k_2,k_3+k_4) 
- \hat I_3^{D+2} (k_1+k_2,k_3,k_4) 
- \hat I_3^{D+2} (k_1,k_2,k_3+k_4)
+ \hat I_4^{D+4}(k_1,k_2,k_3,k_4)
\Bigr].
\nonu
\label{Iscal}
\ear
Now let us compare with our final result \eqref{Gammadecomp}, \eqref{Gammadecompparts}, 
\eqref{gamma},\eqref{hatgamma} for the partially integrated amplitude. We note that in our tensor decomposition
only the following tensors contain terms involving the product $\varepsilon_{12}\varepsilon_{34}$:

\bear
T^{(1)}_{(1234)}, T^{(1)}_{(1243)},T^{(2)}_{(12)(34)}, T^{(4)}_{(12)11}, T^{(4)}_{(12)22},
T^{(4)}_{(34)33}, T^{(4)}_{(34)44}, T^{(5)}_{(12)12}, T^{(5)}_{(12)21},
T^{(5)}_{(34)34}, T^{(5)}_{(34)43}.
\nonu
\ear
Thus separating out these terms and equating their sum to \eqref{GddGdd}, we get the following equation,

\bear
\hat I_{\rm scal} 
&=&
\hat\Gamma^{(1)}_{(1234)}k_{23}k_{14} 
+ \hat\Gamma^{(1)}_{(1243)}k_{24}k_{13} 
+ \hat\Gamma^{(2)}_{(12)(34)}k_{12}k_{34} 
\nonu
&&+\hat\Gamma^{(4)}_{(12)11} \frac{k_{12}k_{13}k_{14}}{k_{34}} 
+ \hat\Gamma^{(4)}_{(12)22} \frac{k_{12}k_{23}k_{24}}{k_{34}}
+\hat\Gamma^{(4)}_{(34)33} \frac{k_{34}k_{31}k_{23}}{k_{12}}
+ \hat\Gamma^{(4)}_{(34)44}\frac{k_{34}k_{41}k_{24}}{k_{12}}
\nonu
&&+ \hat\Gamma^{(5)}_{(12)12} \frac{k_{12}k_{13}k_{24}}{k_{34}} 
+ \hat\Gamma^{(5)}_{(12)21} \frac{k_{12}k_{23}k_{14}}{k_{34}} 
+ \hat\Gamma^{(5)}_{(34)34} \frac{k_{34}k_{31}k_{24}}{k_{12}}
+ \hat\Gamma^{(5)}_{(34)43} \frac{k_{34}k_{41}k_{23}}{k_{12}} \,.
\nonu
\label{idmatchscal}
\ear
The analogous equations for $\ddot G_{13}\ddot G_{24}$ and $\ddot G_{14}\ddot G_{23}$ are obtained from this equation
by the appropriate permutations. The set of these three equations we will call the ``matching identities'' for scalar QED.  

\subsection{Spinor QED}

In the spinor QED case, we do the same procedure starting from the super master formula \eqref{supermaster}. 
Instead of $\ddot G_{12}\ddot G_{34}$ we now have to consider
\bear
D_1D_2\hat G_{12} D_3D_4\hat G_{34} = (\theta_1\theta_2 \ddot G_{12} - G_{F12}) (\theta_3\theta_4 \ddot G_{34} - G_{F34}) \,.
\ear
To saturate the Grassmann integrals, this must be complemented by terms from the exponent. Performing the Grassmann integrals, 
and using $G_{Fij}^2=1$, one finds that $I_{\rm scal}$ of \eqref{Iscal} generalises to 
\bear
\hat I_{\rm spin} &=& 
\hat  I_4^D(k_1,\ldots,k_4) 
\bigl[\ddot G_{12}\ddot G_{34} -\ddot G_{12} k_{34} - \ddot G_{34}k_{12} +k_{12}k_{34} 
+G_{F12}G_{F23}G_{F34}G_{F41} k_{23}k_{14}
\nonu && \hspace{100pt}
+G_{F12}G_{F24}G_{F43}G_{F31} k_{13}k_{24}
\bigr]
\nonu
&=&
\hat I_{\rm scal} 
+ 2(k_{12}+k_{34}) \hat I_4^{D+2}(k_1,k_2,k_3,k_4) 
- 2k_{34} \hat I_3^D(k_1+k_2,k_3,k_4)
- 2k_{12} \hat I_3^D(k_1,k_2,k_3+k_4)
\nonu&&
+ \hat I_4^{D}(k_1,k_2,k_3,k_4)[k_{12}k_{34}
+ k_{23}k_{14} G_{F12}G_{F23}G_{F34}G_{F41}
+ k_{13}k_{24} G_{F12}G_{F24}G_{F43}G_{F31}] \,. \nonu
\label{GddGddspin}
\ear
Thus for spinor QED we get the same matching identity \eqref{idmatchscal} with $\hat I_{\rm scal}$ replaced by $\hat I_{\rm spin}$,
and the coefficient functions $\hat \Gamma^{(i)}_{\cdots}$ replaced by their spinor QED equivalents. 

In part 4 of this series, we will use these matching identities to avoid having to carry through the tensor reduction for the
four-cycle terms, which are by far the most laborious ones in this respect. 

\section{Low-energy limit of the off-shell N-photon amplitudes}\label{lowlimit8}

At the one-loop level, the calculation of amplitudes with large numbers of photons is presently still not feasible, except in the
low-energy limit which we will consider next. In the definition of this limit, care must be taken because of gauge invariance.
In the case of, say, 
the one-loop $N$-point amplitudes in (massive) $\phi^3$ theory, the low-energy limit
would be simply defined by giving all incoming particles their minimal possible energy, $E=m$. This definition is not possible in
the photon case, because nullifying the four-momentum of any photon would mean that the corresponding vertex operator
\eqref{defphotonvertop} turns into the integral of a total derivative,
\bear
V^{\gamma}_{\rm scal}[0,\varepsilon] = \int_0^T d\tau \, \varepsilon\cdot \dot x (\tau) = \varepsilon\cdot (x'-x)\,,
\ear
which vanishes in the closed-loop case. Instead, 
for on-shell photon amplitudes one defines the low-energy limit by the condition that all photon energies
be small compared to the scalar mass, which is the natural mass scale of this amplitude:
\bear
\omega_i \ll m, \quad  i=1,\ldots, N \,.
\ear
For off-shell photons, one has to separately require that also 
\bear
\abs{{\bf k}_i}\ll m, \quad  i=1,\ldots, N \,.
\ear
These conditions then justify truncating all the vertex operators to their terms linear in the momentum. Thus
we define the vertex operator of a low-energy photon by

\begin{figure}[h]
		\centering 
		\includegraphics[width=.15\textwidth]{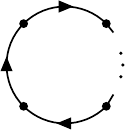}
		\caption{One-loop $n$ - point worldline Feynman diagram.}
		\label{srednicki}
	\end{figure}

\bear
V^{\gamma\, ({\rm LE})}_{\rm scal} [k,\varepsilon]
&\equiv & \int_0^Td\tau \, \varepsilon\cdot \dot x(\tau)\, i k\cdot x(\tau)\,,
\label{defVgammaLE}
\ear
(it is this factor of momentum for each photon that is responsable for the steep $\omega^6$ falloff of the photon-photon cross section 
in the low-energy limit, mentioned in the introduction).
Adding a total derivative term, we can rewrite this vertex operator in terms of the photon field strength tensor:

\bear
V^{\gamma\, ({\rm LE})}_{\rm scal} [f] &\equiv & 
V^{\gamma\, ({\rm LE})}_{\rm scal} [k,\varepsilon] - \frac{i}{2} 
\int_0^Td\tau \,\frac{d}{d\tau} \bigl(\varepsilon\cdot x(\tau)\, k\cdot x(\tau)\bigr)
=
\frac{i}{2} \int_0^Td\tau \, x(\tau)\cdot f \cdot \dot x(\tau) \, .
\label{defVgammaLEfin}
\nonumber\\
\ear
The Wick contraction of a product of such objects consists of terms that are
products of Lorentz cycles $\tr (f_{i_1}f_{i_2}\cdots f_{i_n})$ with coefficients that, by suitable IBPs, 
can be written as integrals of $\tau$ - cycles $\dot G_{i_1i_2}\dot G_{i_2i_3} \cdots \dot G_{i_ni_1}$.  
The result can be further simplified by observing that, in the one-dimensional worldline QFT, each factor of

$$\int_0^T d\tau_{i_1} \cdots \int_0^Td\tau_{i_n} \dot G_{i_1i_2}\dot G_{i_2i_3} \cdots \dot G_{i_ni_1}\tr (f_{i_1}f_{i_2}\cdots f_{i_n})\,,$$ 

\noindent
can be identified with the one-loop $n$ - point Feynman diagram depicted in Fig. \ref{srednicki}.

Products of such factors in the worldline theory correspond to 
disconnected diagrams, and by standard combinatorics can be reduced to the exponential of the
sum of all connected diagrams. In this way, and with the usual rescaling $\tau_i = Tu_i$, we arrive at 

\bear
\bigl\langle
V^{\gamma\, ({\rm LE})}_{\rm scal} [f_1] 
V^{\gamma\, ({\rm LE})}_{\rm scal} [f_2] 
\cdots
V^{\gamma\, ({\rm LE})}_{\rm scal} [f_N] 
\bigr\rangle
=
i^N T^N
\,\exp\biggl\lbrace \sum_{m=1}^{\infty}b_{2m}
\sum_{\lbrace i_1\ldots i_{2m}\rbrace}
Z^{\rm dist}_{2m}(\lbrace i_1i_2\ldots i_{2m}\rbrace)\biggr\rbrace
\bigg \vert_{f_1\ldots f_N}\,,
\nonumber\\
\label{WickLE}
\ear
where 
$Z^{\rm dist}_k(\lbrace i_1i_2\ldots i_k\rbrace)$
denotes the sum over all distinct
Lorentz cycles which can be formed with a given
subset of indices, e.g. 
$Z^{\rm dist}_4(\lbrace ijkl\rbrace) = Z_4(ijkl) + Z_4(ijlk) + Z_4(ikjl)$,
and $b_n$ denotes the basic ``bosonic cycle integral''
\bear
b_n &\equiv& \int_0^1 du_1du_2\ldots du_n\,
\dot G_{12}\dot G_{23}\cdots\dot G_{n1} \,.
\label{chainintbos}
\ear
This integral can be expressed in terms of the Bernoulli numbers ${\cal B}_n$ \cite{Schmidt1993-438}
\bear
b_n 
 =
\qquad\left\{ \begin{array}{r@{\quad\quad}l}
-2^n{{\cal B}_n\over n!}  & \qquad n{\rm \quad even}\,,\\
0 & \qquad n{\rm \quad odd}\,.\\
\end{array} \right.
\label{10-bn}
\ear
Eq. \eqref{WickLE} can be simplified using the combinatorial fact that
\bear
{\rm tr}\Bigl[(f_1+\ldots +f_N)^n\Bigr]\bigg\vert_{\rm all\,\,different}
&=&
2n \sum_{\lbrace i_1\ldots i_n\rbrace} Z^{\rm dist}_n(\lbrace i_1i_2\ldots i_n\rbrace)\,,
\nonumber\\
\label{F=F}
\ear
($n\ne 0$). Introducing $f_{\rm tot} \equiv \sum_{i=1}^N f_i$, 
using all this in \eqref{Nphotonvertop}
and eliminating the $T$-integral,
we arrive at  the following formula for the low-energy limit of the one-loop
$N$-photon amplitude \cite{Dunne_2002}:
\bear
\Gamma_{\rm scal}^{({\rm LE})}
(k_1,\varepsilon_1;\ldots ;k_N,\varepsilon_N)
&=&
\frac{e^N \Gamma(N-\frac{D}{2})}{(4\pi)^2m^{2N-D}}
\,\exp\biggl\lbrace \sum_{m=1}^{\infty}\frac{b_{2m}}{4m} \tr (f_{\rm tot}^{2m})
\biggr\rbrace
\bigg \vert_{f_1\ldots f_N} \,.
\nonumber\\
\label{Nphotlowfin}
\ear
According to the above, the transition from scalar to spinor QED at this level can be effected by replacing the
``chain integral'' (\ref{chainintbos}) by the ``super chain integral'' 
\bear
\int_0^1 du_1du_2\ldots du_n\,
\Bigl(\dot G_{12}\dot G_{23}\cdots\dot G_{n1} 
-
G_{F12}G_{F23}\cdots G_{Fn1}\Bigr)
=
(2-2^n)\,b_n \,. \nonumber\\
\label{chainintferm}
\ear
The only other change is a global factor of $(-2)$ for statistics and degrees of freedom. Therefore
\bear
\Gamma_{\rm spin}^{({\rm LE})}
(k_1,\varepsilon_1;\ldots ;k_N,\varepsilon_N)
&=& -2
\frac{e^N \Gamma(N-\frac{D}{2})}{(4\pi)^2m^{2N-d}}
\,\exp\biggl\lbrace \sum_{m=1}^{\infty}(1-2^{2m-1})\frac{b_{2m}}{2m} \tr (f_{\rm tot}^{2m})
\biggr\rbrace
\bigg \vert_{f_1\ldots f_N} \,.
\nonumber\\
\label{Nphotlowfinspin}
\ear
In the four-point case this gives (setting now also $D=4$)

	\bear
	\Gamma_{\rm scal}(k_1,\varepsilon_1 ;... ;k_4,\varepsilon_4 )&=&\frac{e^ 4}{(4\pi)^2 m^4}\,  
	\Big\{b_4 [ Z_4(1234)+Z_4(1243)+Z_4(1324)]\nonumber\\
	&& + b_2^2[ Z_2(12)Z_2(34)+ Z_2(13)Z_2(24) + Z_2(14)Z_2(23)]\Big\}\,, \\
	\Gamma_{\rm spin}(k_1,\varepsilon_1 ;... ;k_4,\varepsilon_4 ) &=& -\frac{2e^ 4}{(4\pi)^2 m^4}
	\, \Big\{-14  b_4  \left[ Z_4(1234) + Z_4(1243) + Z_4(1324)\right]\nonumber \\
	&&+ (-2 b_2)^2 \left[ Z_2(12) Z_2(34) + Z_2(13)Z_2(24) + Z_2(14)Z_2(23)\right] \Big\}\,.
	\nonumber\\
	\ear
	
These expressions can also be derived from the effective Lagrangians \eqref{EH} resp. \eqref{weisskopf},
see \cite{Itzykson1980rh,Berestetskii1982-501,Martin2003-335}.
	
\section{Computational strategy}

Our goal in this series of paper is to calculate the four-photon amplitudes fully off-shell, and with any number of
the photons taken in the low-energy limit. In the previous chapter we have already settled the case where all photons
are low-energy, and by momentum conservation it would not make sense to take three of them low-energy, and the fourth not.
This leaves the cases of zero, one or two low-energy photons. Let us thus discuss our general strategy for treating
each of these cases starting from the worldline representation. It consists of three elements:

\benn

\item
{\it Integrating out a low-energy leg:} when a photon leg is taken to be at low energy, the corresponding $u_i$ - integral becomes
polynomial and can be integrated out. For this purpose, it is not necessary to write the integrand out explicitly, since there are
formulas available \cite{Schubert2001-73} that will allow us to perform all the integrals that can appear at the four-point level, with the result
already written in terms of the worldline Green's functions between the remaining points.  
Some examples are

\begin{equation}\label{int1}
\int_0^1 du_1 G_{12} = \frac{1}{6}\,,
\end{equation}
\begin{equation}\label{int2}
\int_0^1 du_1 \dot{G}_{12} = 0\,,
\end{equation}
\begin{equation}\label{int3}
\int_0^1 du_3 G_{13} G_{32} = -\frac{1}{6} G_{12}^2 + \frac{1}{30}\,,
\end{equation}
\begin{equation}\label{int4}
\int_0^1 du_3 \dot{G}_{13} G_{32} = -\frac{1}{3}\dot{G}_{12}G_{12}\,,
\end{equation}
\begin{equation}\label{int5}
\int_0^1 du_3 \dot{G}_{13} \dot{G}_{32} = 2 G_{12} -\frac{1}{3}\,,
\end{equation}
\begin{equation}\label{int6}
\int_0^1 du_4 \dot{G}_{41} \dot{G}_{42} \dot{G}_{43} = - \frac{1}{6}(\dot{G}_{12}-\dot{G}_{23})(\dot{G}_{23}-\dot{G}_{31})(\dot{G}_{31}-\dot{G}_{12})\,.
\end{equation}

\item
{\it Tensor reduction by IBP:} as usual, we are aiming at a reduction to purely scalar integrals, which in the worldline
representation means that we wish to remove the polynomial prefactors of the universal exponential $\e^{(\cdot)}$. 
Although at the one-loop level there are various ready-made tensor reduction schemes available, to use those we would
have to first split the worldline parameter integrals into the ordered sectors to convert them into Feynman-Schwinger parameter integrals.
Here, we will rather develop tensor reduction schemes that are adapted to the worldline representation and make it possible
to perform the reduction to scalar integrals without the need to fix an ordering for the photon legs. 

\item
{\it Tensor reduction by differentiation:} any factor of $\dot G_{ij}^2$ in the integrand can, using the identity
$\dot{G}_{ij}^2 = 1 - 4 G_{ij}$, be generated from the universal exponential through a derivative with respect
to $k_i\cdot k_j$. 

\enn

The ``two-low'' case will involve only integration, while in the ``one-low'' case integration and tensor
reduction will have to be combined. For ``zero low'' we have a pure tensor reduction task.

The final (dimensionally regularized) scalar integrals will be expressible in terms of hypergeometric functions, namely 
$_2F_1$ in the two-low case (as for two-point functions), 
$_2F_1$ and $F_1$ in the one-low case (as for three-point functions), 
and $_2F_1$, $F_1$ and the Lauricella-Saran function for general kinematics (see \cite{Riemann:2019gqf} and refs. therein). 
%
%

\section{Summary and outlook}

In this first part of a series of four papers on the (scalar and spinor) QED four-photon amplitudes fully off-shell,
our main purpose was to motivate this whole effort, and to prepare the ground for the following parts, where these
amplitudes are calculated with general kinematics (part 4), and with one (part 3) or two (part 2) legs taken in the
low-energy limit. Thus these sequel papers are all meant to be used together with the present part 1 (although a reader
of part 4, for example, would not need to look at part 2 or part 3, if his or her interest is only in the case of general kinematics).  
The simplest case where all four photons are low energy is a textbook calculation but has been included here for completeness, 
and as a warm-up demonstrating the efficiency of the formalism which allows one to trivialize this computation even for an arbitrary number of photons. 

The main new result of part 1 is a further refinement of the worldline IBP procedure, originally proposed by Bern and Kosower
in the gluonic context, and later systematized by Strassler. It leads to an extremely compact integrand for the
four-photon amplitudes in both scalar and spinor QED, and at the same time projects these amplitudes, already at the integrand
level, into the well-known minimal transversal basis of five tensors proposed by Costantini et al. in 1971.  
Since the four-photon amplitude is the prototype of all amplitudes with four gauge bosons, we expect this to lead to similar improvements for
other cases, such as the form-factor decomposition
of the four-gluon amplitudes obtained in \cite{Ahmadiniaz:2016-1642004} using the same formalism.

We have outlined the strategy that we will follow in the calculation of the integrals for the various cases, including the details of 
the integrating out of a low-energy photon 
\footnote{In the very recent \cite{Ludtke:2023hvz} a similar strategy was followed in the dispersive calculation of the hadronic light-by-light contribution to the muon anomalous magnetic momentum.}, 
and derived a set of ``matching identities'' that will be used in part 4 to facilitate the tensor reduction, but which we include here since they may well
turn out to be of more general significance.

Further, we have reviewed the history of photonic processes in general and the four-photon amplitudes in particular, with an
emphasis on processes that naturally involve some off-shell photons, either because external fields are involved or we use the amplitude
as a building block for a higher-order process. We have also given a summary of the worldline approach to the calculation
of such processes. 

The improved worldline representation of the four-photon amplitudes developed here 
offers significant advantages also for the on-shell case.
A complete recalculation of the massive scalar and spinor QED on-shell four-photon amplitudes along the present lines is in progress
and will be presented in a separate publication.

\bigskip

\no
{\bf Acknowledgements:} We thank D. Bernard, G. Colangelo, A. Davydychev, 
H. Gies, F. Karbstein, T. Riemann, K. Scharnhorst and R. Shaisultanov for helpful
conversations or correspondence. 
M. A. Lopez-Lopez is grateful to CoReLS and C.H. Nam for his hospitality during the first stages of this work.
C. Lopez-Arcos, M. A. Lopez-Lopez and C. Schubert thank CONACYT for financial support. 

\appendix
\renewcommand*{\thesection}{\Alph{section}}

\section{Conventions}
\label{app1}

We use natural units $c = \hbar = 1$.
We work in euclidean space with a positive definite metric $(g_{\mu\nu})={\rm diag}(++++)$. 
Minkowski space amplitudes with metric $(\eta_{\mu\nu})={\rm diag}(-+++)$ are obtained by analytically continuing 
\bear
g_{\mu\nu}&\rightarrow& \eta_{\mu\nu}\,,\non
k^4&\rightarrow&-i k^0\,,\non
T&\rightarrow& is\, .
\ear
We use 
the absolute value of the electron charge $e = |e|$, corresponding to a covariant derivative $D_{\mu} = \partial_{\mu} + ieA_{\mu}$. 
Momenta of external photons are ingoing. 

%
%
%


\bibliography{LbL}

\end{document}